\documentclass[aps,english,prb,reprint,superscriptaddress,showpacs]{revtex4-1}

\usepackage{graphicx}
\usepackage{amssymb}
\usepackage{babel}
\usepackage{color}
\usepackage{multirow}
\usepackage[T1]{fontenc}
\usepackage{subfigure}

\begin{document}
\title{\textbf{Exploring the equilibrium and dynamic phase transition properties of Ising 
ferromagnet on a decorated triangular lattice}}

\author{Y. Y\"{u}ksel}
\affiliation{Dokuz Eylul University, Faculty of Science, Physics Department, 
Tinaztepe Campus, 35390 Izmir, Turkey}

\date{\today}

\begin{abstract}
We study the equilibrium and dynamic phase transition properties of two-dimensional Ising model on a  decorated triangular lattice under the influence of a time-dependent magnetic field composed 
of a periodic square wave part plus a time independent bias term. Using Monte Carlo simulations with standard Metropolis algorithm, we determine the equilibrium critical behavior in zero field. 
At a fixed temperature corresponding to the multidroplet regime, we locate the relaxation time and the dynamic critical half-period at which a dynamic phase transition 
takes place between ferromagnetic and paramagnetic states.  
Benefiting from finite-size scaling theory, we estimate the dynamic critical exponent ratios for the dynamic order parameter and its scaled variance, respectively. The response function of the average energy is
found to follow a logarithmic scaling as a function of lattice size. At the critical half-period and in the vicinity of small bias field regime, average of the dynamic order parameter 
obeys a scaling relation with a dynamic scaling exponent which is very close to the equilibrium critical isotherm value. Finally, in the slow critical dynamics regime,
investigation of metamagnetic fluctuations in the presence of bias field 
revels a symmetric double-peak behavior for the scaled variance contours of dynamic order parameter and average energy. Our results strongly resemble those previously reported for kinetic Ising models.
\end{abstract} 

\maketitle

\section{Introduction}\label{intro}
Kinetic Ising model and its variants \cite{tome,lo,vatansever_3} which focus on the response of a ferromagnetic (FM) system to a time-varying and externally applied periodic magnetic field $h(t)$ 
have been a class of the most actively studied   
problems of statistical mechanics, as well as in the theory of phase transitions and critical phenomena. 
In this model, a dynamic phase transition may originate as a result of a dynamic symmetry breaking mechanism \cite{chakrabarti,riego_2}. 
This mechanism depends on a competition between two characteristic time scales; namely the period $P$ of the oscillating magnetic field, and the relaxation time $\tau$ of the system. Despite the fact that the 
period $P$ is an adjustable external parameter, $\tau$ depends on several factors including the temperature, field amplitude, and other magnetic interaction parameters such as the ferromagnetic exchange coupling which 
mimics the interaction between neighboring magnetic moments in the lattice. If $\tau$ is larger than period $P$, the system cannot find enough time to follow the external perturbation, hence the instantaneous magnetization
$m(t)$ oscillates around some non-zero value in which the magnetic phase of system is called ``\textit{dynamic ferromagnetic}''. On the contrary, when $P >\tau$, $m(t)$ can easily follow 
the alternation of magnetic field with some small delay. In this case, the system is in the ``\textit{dynamic paramagnetic}'' phase. 

In addition to the theoretical observations, dynamic phase transition (DPT)
properties of magnetic systems have been experimentally realized in some recent works \cite{berger,robb_2} in which some similarities have been unveiled between DPT and its equilibrium counterpart called the 
thermodynamic phase transitions (TPT). For instance, it was shown that a time-independent bias field $h_{b}$ in DPT plays the role of the homogeneous magnetic field in TPT. Therefore, bias field $h_{b}$ is identified
as the ``conjugate field'' of the dynamic order parameter (i.e. the period-averaged magnetization) $\langle Q\rangle$  \cite{robb_1}. 
In the presence of an oscillating magnetic field with square wave form in addition to the bias field $h_{b}$, $(\langle Q\rangle-h_{b})$ curves obtained in the vicinity of dynamical 
critical point were found to show a power law behavior with a 
dynamic scaling exponent $\delta_{d}$ which is identical to the critical isotherm $\delta_{e}$ of TPT \cite{robb_1}.
Furthermore, universality and scaling relations in DPT have been examined in two- \cite{sides_1,korniss,buendia} and three- \cite{park} dimensions, and some additional similarities
between DPT and TPT cases were reported in the absence of bias field. A number of general outcomes can be summarized as follows: 
a DPT can be observed in the vicinity of critical period $P_{c}$ below which a dynamically ordered phase is manifested.
In this regard, $\langle Q\rangle$ versus $P$ curves of DPT qualitatively exhibit the same behavior as the spontaneous magnetization versus temperature curve of TPT. 
It is worth to note that the critical period $P_{c}$ at which a DPT takes place between dynamically ordered and disordered states was found to be highly sensitive to the field amplitude \cite{korniss}.
Apart from these, the most striking outcome is that the universality class of DPT is the same as the corresponding TPT in the vicinity of $P_{c}$. This latter result is also 
found to be robust against introduction of quenched disorder \cite{vatansever_e2}.
For a detailed discussion of the scaling properties and phase diagrams of DPT in low dimensional, semi-infinite, and bulk systems, please refer to Ref. \cite{yuksel1}. 
However, these similarities between DPT and TPT cases should be approached with utmost caution, since in the presence of bias field some features of DPT substantially differ from those observed in TPT. For instance,
for a regular ferromagnet in the presence of longitudinal magnetic field, magnetic susceptibility curve as a function of magnetic 
field exhibits a broad symmetric maximum which is centered around zero field \cite{berger_eq} whereas in the DPT counterpart, magnetic susceptibility (as well as scaled variance) plotted against 
bias field exhibit multiple symmetric peaks which are called ``\textit{meta-magnetic anomalies}'' \cite{riego}. 

After the discovery of graphene as a two-dimensional (2D) material \cite{novoselov,geim}, there has been a renowned interest in 2D magnetism during the last decade. 
Consequently, from the view-point of dynamic phase transition phenomenon, investigation of 2D lattices gained particular importance. However, it is  worth mentioning that the vast majority 
of the literature on the kinetic Ising model discussed so far is restricted to regular lattices, and in general, the role of non-regular lattices has been overlooked.
In the present work, in order to overcome this issue, we perform extensive Monte Carlo simulations on a decorated triangular lattice (DTL) to estimate both the TPT and DPT characteristics of this non-regular lattice.
To the best of our knowledge, thermal and magnetic properties of a DTL was scarcely investigated before. Among these works, one can refer to Refs. \cite{jabar,galisova} for ferri-magnetic and magnetocaloric properties, and to Ref. \cite{azhari} for 
treatment of Blume-Capel model. Note that the main focus of these works is limited to TPT case. Therefore, the objective of the present paper is to provide a detailed analyses of 
the equilibrium critical behavior of TPT, as well as
the critical exponents corresponding to DPT case in the presence of a time dependent square magnetic field with period $P$ for a kinetic Ising model located on a DTL. 
In addition to these properties, we also clarify the meta-magnetic anomalies
and power-law behavior in $(\langle Q\rangle-h_{b})$ curves with a DPT scaling exponent $\delta_{d}$. 

The outline of the paper is as follows: In Sec. \ref{formulation}, we briefly introduce our model and simulation details. Sec. \ref{results} contains our simulation results and related discussions. 
Finally, Sec. \ref{conclude} is devoted for the concluding remarks.

\section{Model and Formulation}\label{formulation}
\begin{figure}[!h]
\center
\includegraphics[width=5.5cm]{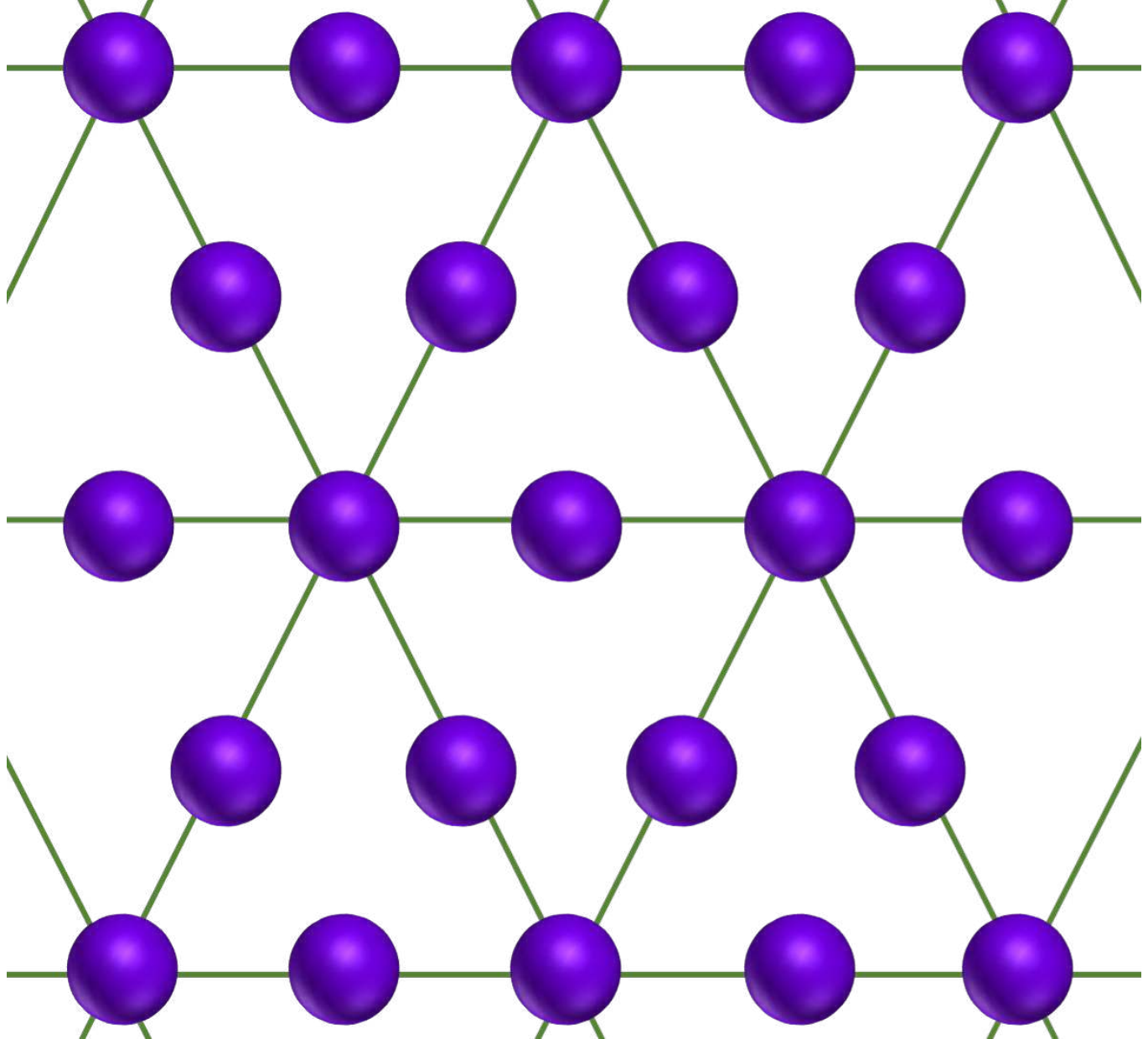}
\caption{Schematic representation of decorated triangular lattice. }\label{fig1}
\end{figure}

We simulate the system defined by the Hamiltonian 
\begin{equation}\label{eq1}
\mathcal{H}=-J\sum_{\langle ij\rangle}S_{i}^{z}S_{j}^{z}-\sum_{i}h(t)S_{i}^{z}, 
\end{equation}
where $J>0$ is the ferromagnetic exchange coupling between nearest-neighbor spins and $S_{i}^{z}$ is a pseudo spin variable taking the values $\pm1$. Each spin is located on the nodes of a DTL
which is schematically represented in Fig. \ref{fig1}. The last term in Eq. (\ref{eq1}) stands for the Zeeman term where the magnetic field $h(t)$ is composed of two parts as a time independent bias term $h_{b}$
and a time dependent part in square-wave form. We implement Monte Carlo simulations based on standard Metropolis algorithm \cite{binder} by imposing periodic boundary conditions (PBC) applied in each direction. 
The lattice sites are swept randomly and one Monte Carlo step (MCS) consists of $N=L\times L$ spin-flip attempts where $L$ is the linear dimension of DTL depicted in Fig. \ref{fig1}.

\subsection{Measured quantities and simulation parameters for TPT properties}
In order to clarify the equilibrium critical behavior, we set field amplitude and bias field terms to zero $(h_{0}/J,h_{b}/J=0)$ and measure the following quantities by considering $100$ individual samples and 
$5\times 10^{4}$ Monte Carlo steps at each temperature after discarding the first $20\%$ for thermalization:
\begin{itemize}
 \item Spontaneous magnetization:
 \begin{equation}\label{eq2}
 M=\frac{1}{N}\left\langle \sum_{i=1}^{N}S_{i}^{z}\right\rangle, 
 \end{equation}
\item Magnetic susceptibility:
\begin{equation}\label{eq3}
\chi=N\{\langle M^{2}\rangle-\langle M\rangle^{2}\}/k_{B}T,
 \end{equation}
 \item Internal energy and specific heat:
 \begin{equation}\label{eq4}
 \langle E \rangle=\langle \mathcal{H}\rangle, \quad \mathcal{C}=\frac{\partial \langle E \rangle}{\partial T}, 
 \end{equation}
\end{itemize}
where angular brackets denote thermal averaging. Note that we set $k_{B}=1$ for simplicity.

\subsection{Measured quantities and simulation parameters for DPT properties}
To determine the DPT properties, we perform a series of simulations by considering lattice sizes ranging between $64\leq L \leq 324$. In DPT case, the time length of a simulation
for a given set of parameters with a fixed $L$ depends on the period $P$ of the periodic magnetic field. In this regard, to calculate the physical quantities, $2.2\times 10^{4}$ period 
cycles of the oscillating field was considered and the initial 
$2\times10^{3}$ cycles were discarded for thermalization. In the absence of bias field, we take $500$ independent realizations to reduce the statistical errors. This number of samples were found to be sufficient 
to obtain high quality data around the dynamic critical point $P_{c}$ to estimate the critical exponent ratios. 
Accumulated running averages calculated around $P_{c}$ have been displayed in Appendix (c.f. see Fig. \ref{fig10}). 
Moreover, in order to perform error analysis, we use Jackknife method \cite{newman} and 
to estimate the error bars, we divide the data set containing 500 individual 
measurements for each quantity into 20 subgroups. Note that the size of 
the obtained error bars are genrally smaller than the size of the data points.

Once we monitor the time series $m(t)$ of instantaneous magnetization, it is possible to define the dynamic order parameter $Q(k)$ at the $k^{th}$ cycle of the dynamic magnetic field 
\begin{equation}\label{eq5}
Q(k)=\frac{1}{(2t_{1/2})}\int_{(k-1)(2t_{1/2})}^{k(2t_{1/2})} m(t)dt,
\end{equation}
where we prefer to use the parameter $t_{1/2}$ for convention which defines the half-period of the dynamic magnetic field, i.e. we set $P=2 t_{1/2}$.
Using Eq. (\ref{eq5}), we calculate the dynamic order parameter $\langle Q\rangle$ which is the average of $Q(k)$ where the averaging is performed over many cycles of $h(t)$. In addition, 
dynamic scaling variance of $Q$
which resembles the dynamic magnetic susceptibility \cite{robb_1} is given by the formula
\begin{equation}\label{eq6}
\chi_{Q}=N\left[\langle Q^{2}\rangle_{L}-\langle Q\rangle^{2}_{L} \right].
\end{equation}
Following the same procedure, dynamic scaling variance of average internal energy can also be obtained from
\begin{equation}\label{eq7}
\chi_{E}=N\left[\langle E^{2}\rangle_{L}-\langle E\rangle^{2}_{L} \right],
\end{equation}
where $\langle E\rangle$ is the average internal energy per spin calculated using the Hamiltonian (\ref{eq1}).
Last but not least, we also measure the Binder cumulant $V_{L}$
\begin{equation}\label{eq8}
V_{L}=1-\frac{\langle Q^{4}\rangle}{3\langle Q^{2}\rangle}, 
\end{equation}
benefiting from the higher order moments of $Q$ to precisely determine the critical point \cite{binder0}.

It is important to underline that DPT takes place in the multidroplet (MD) regime in which the metastable decay originates via nucleation and growth processes of many droplets \cite{sides_1,rikvold_1}.
Therefore, in order to ensure that the system is in the MD regime,
the field amplitude and the temperature are respectively  fixed as $h_{0}/J=0.3$ and $T=0.8T_{c}$ 
throughout the simulations where $T_{c}$ is the pseudo-critical temperature of the DTL.

\section{Results and Discussion}\label{results}
\begin{figure}[!h]
\center
\subfigure[\hspace{0cm}] {\includegraphics[width=4.1cm]{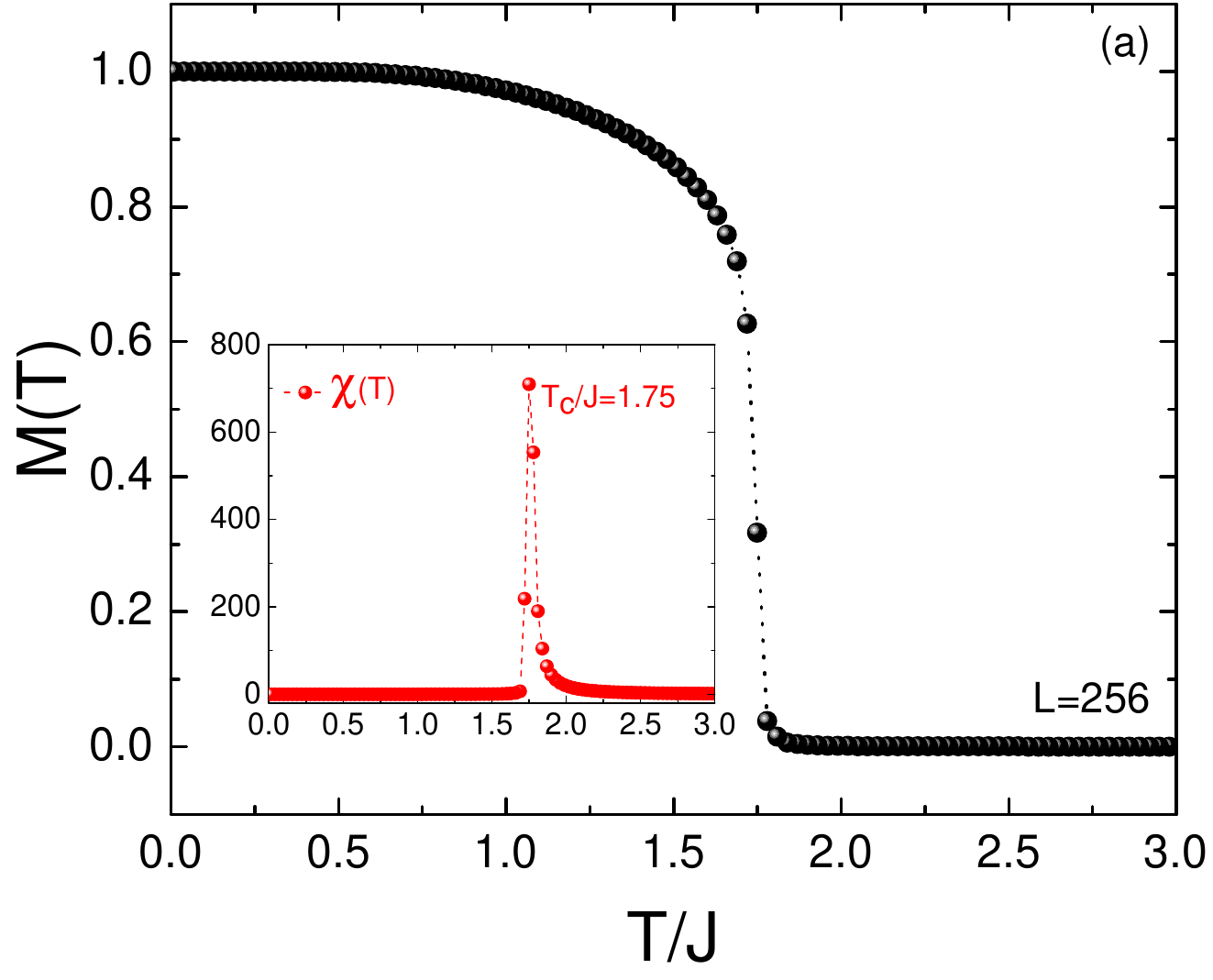}}
\subfigure[\hspace{0cm}] {\includegraphics[width=4.1cm]{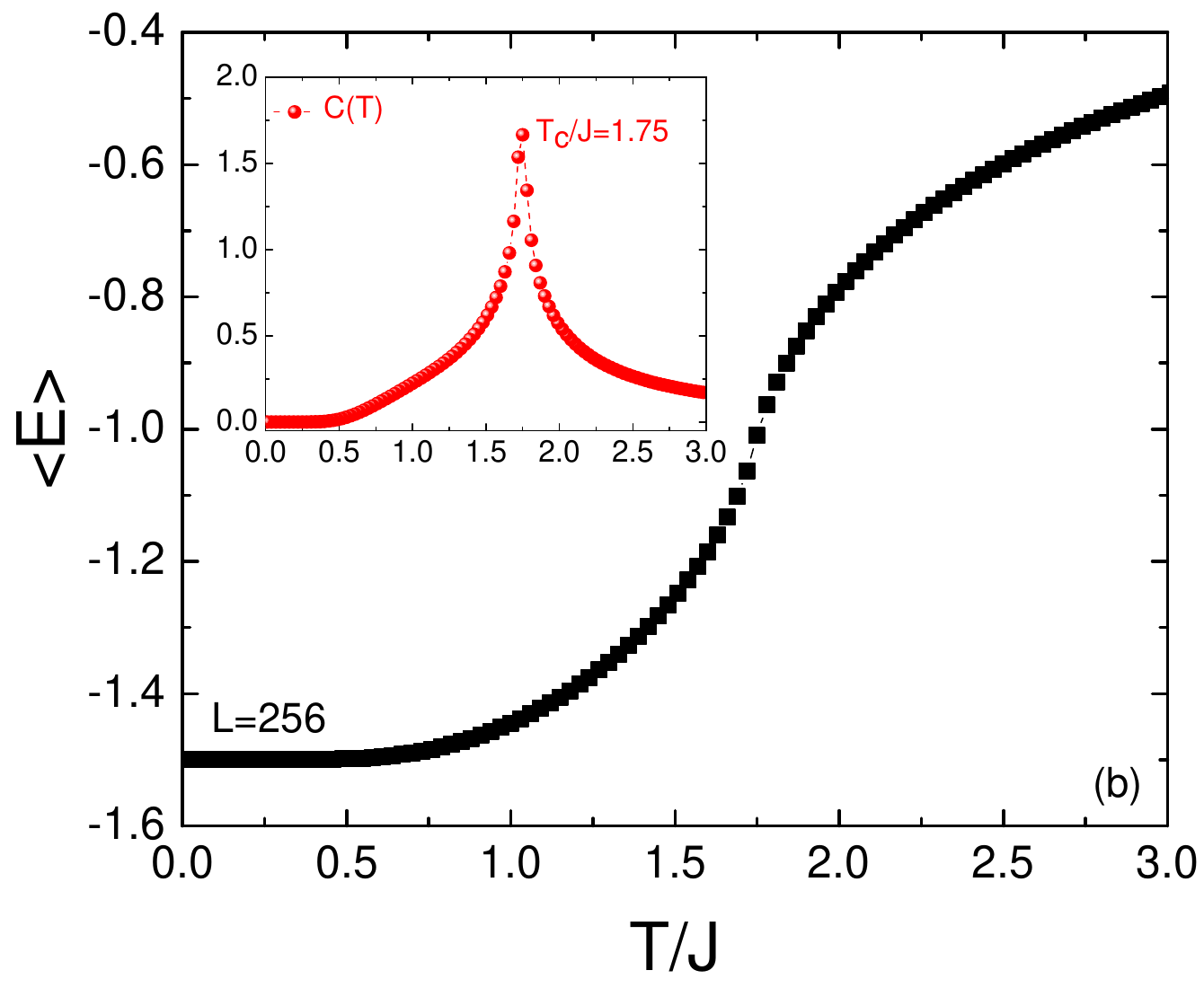}}\\
\caption{Temperature dependencies of thermal and magnetic properties calculated 
at zero field: 
(a) magnetization and magnetic susceptibility, (b) Average energy and heat 
capacity. Each curve is obtained for a lattice with $L=256$.}\label{fig2}
\end{figure}

In order to investigate the equilibrium critical behavior of the system (i.e. TPT case), 
we have calculated the temperature dependencies of thermal and magnetic properties defined by Eqs. (\ref{eq2}-\ref{eq4}) in the absence of magnetic field. 
The results are shown in  Fig. \ref{fig2} for a DTL with $L=256$. 
Spontaneous magnetization of the system depicted in Fig. \ref{fig2}a shows that a ferromagnetic-paramagnetic phase transition emerges at the critical point $T_{c}$, and the transition is of second-order.  
For a DTL, a ratio $3/4$ of lattice sites are coordinated to two nearest-neighbors $(z_{1}=2)$ and the remaining $1/4$ of spins have six nearest-neighbors $(z_{2}=6)$, 
indicating that the effective coordination number of DTL is $Z_{eff}=3$ \cite{azhari}.
Temperature dependent internal energy per spin curve (Fig. \ref{fig2}b) attains a ground state value $\langle E\rangle/J=-1.5$  which eventually supports Ref. \cite{azhari}. 
The insets of Fig. \ref{fig2}a and \ref{fig2}b show the variation of response functions, i.e. the magnetic susceptibility $\chi$ and specific heat $\mathcal{C}$ exhibiting sharp peaks at the critical temperature $T_{c}$. 
Examination of $\chi(T)$ and $\mathcal{C}(T)$ curves reveals that the ordering temperature is $T_{c}/J=1.75$.

Note that the obtained critical temperature value is different than those obtained in Refs. \cite{azhari,vatansever_e2,huan}. The reason is two-fold: In Ref. \cite{azhari}, a spin-1 Blume-Capel model is considered.
Consequently, due to the reduced anisotropy in comparison with the Ising counterpart discussed in the present work, critical temperature is expected to be 
smaller than our numerical result. On the other hand, Refs. \cite{vatansever_e2,huan}
investigate the model on a regular triangular lattice in which the coordination number $Z=6$ is as twice as larger than that of a DTL with $Z_{eff}=3$. Besides, one can also compare our result with $T_{c}/J=1.519$ 
of a honeycomb lattice with $Z=3$ \cite{fisher}. It should be mentioned that although the Binder cumulant analysis give more precise values
for the exact location of $T_{c}$, we do not need to find the  location of the critical temperature in full precision, as our estimated value ensures that the system stays in the MD regime in the presence of a dynamic magnetic field. 
Therefore, in order to reduce the computational time, we benefit from the pseudo critical temperature obtained by inspecting the response functions corresponding to $L=256$ in the following analyses of DPT properties.

\begin{figure}[!h]
\center
\includegraphics[width=7.0cm]{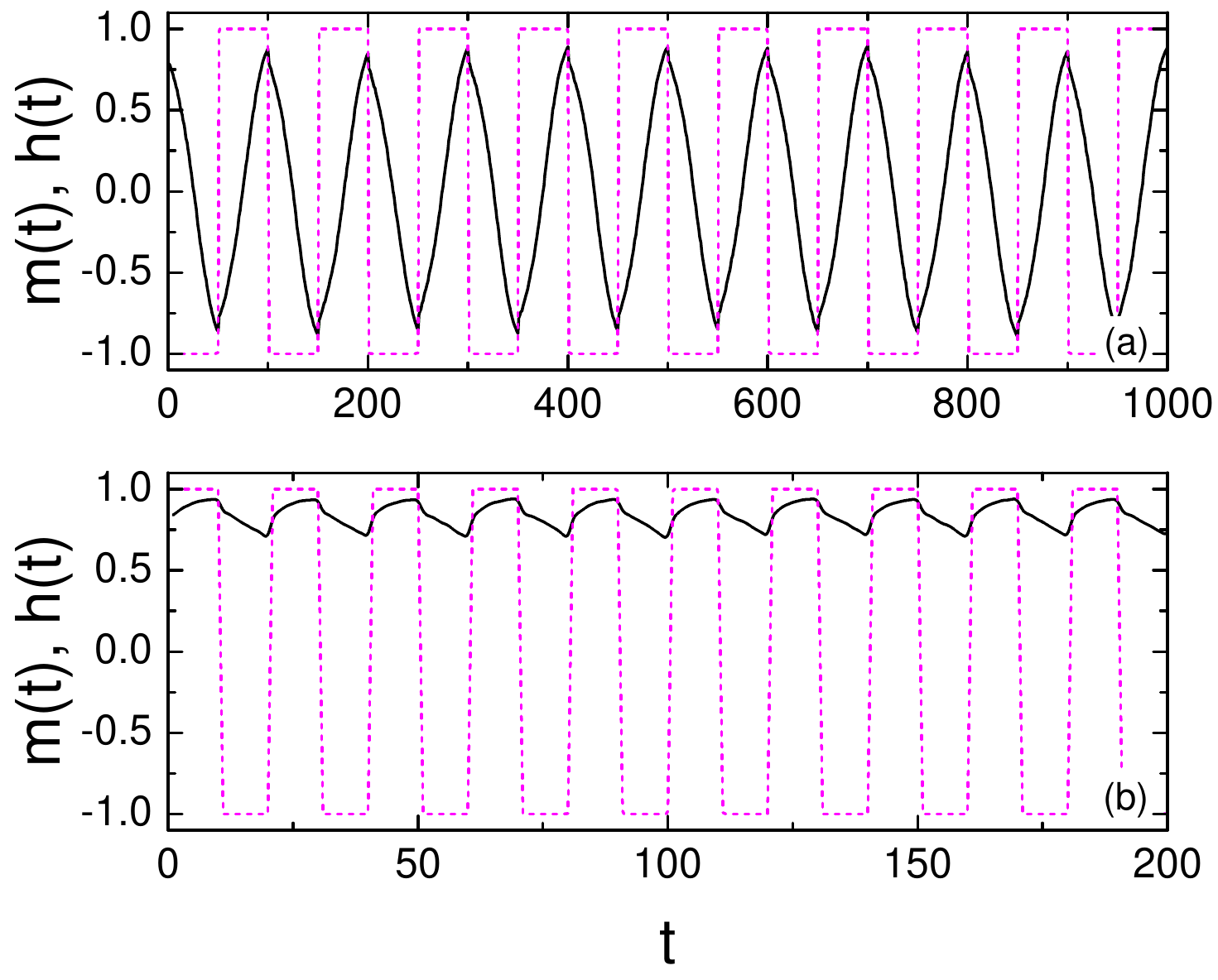}
\caption{Time series of magnetization in the presence of a square-wave magnetic 
field with half period (a) $t_{1/2}=50$, (b) $t_{1/2}=10$. 
The dashed lines represent the magnetic field. In (a) dynamic paramagnetic 
(disordered) regime is manifested whereas in (b) the system exhibits dynamic 
ferromagnetic (ordered) behavior.}\label{fig3}
\end{figure}
Once the critical temperature of the system is determined, we can go one step forward in our analyses of DPT properties. The competition mechanism leading to the emergence of DPT is illustrated in Fig. \ref{fig3}.
When the magnetization is aligned with the magnetic field then  the energy is minimized, and a change in the sign of magnetic field causes the magnetization to flip along the field direction within a certain amount of time. 
If the relaxation time $\tau$ needed to flip the sign of the magnetization in the metastable state is comparable to or smaller than the critical half-period $t_{1/2}^{c}$, 
then a domain nucleation process takes place which is followed by the formation of new domains composed of parallel spins 
along the new field direction. In this case, the magnetization can follow the periodic alternation of the dynamic magnetic field (Fig. \ref{fig3}a) with a small phase lag (dynamic paramagnetic state). 
On the contrary, if $\tau$ is larger than $t_{1/2}^{c}$, the system always stays in the metastable- state indicating that the dynamically ordered state is favored in which the net magnetization oscillates 
around some non-zero value, as shown in Fig. \ref{fig3}b. 

\begin{figure}[!h]
\center
\includegraphics[width=7.0cm]{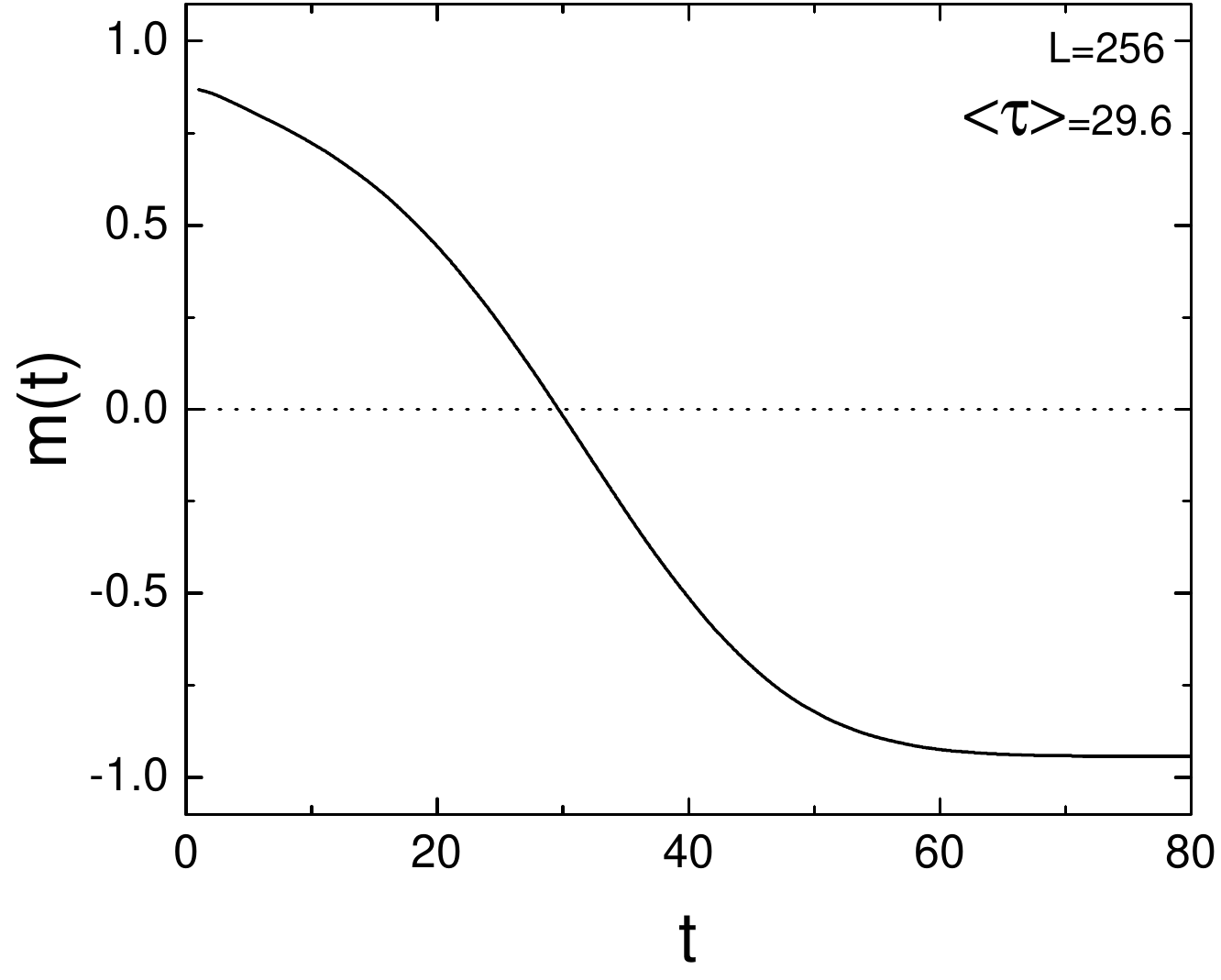}
\caption{Instantaneous magnetization as a function of time calculated at a 
temperature $T=0.8T_{c}$ and a constant bias field $h_{b}/J=0.3$ for a lattice 
with $L=256$. 
The relaxation time (i.e. the metastable life-time) of the system is determined 
by inspecting the crossing point of the curve at which the magnetization  
switches its sign. 
The dashed horizontal line refers to a guide to the eye of the 
reader.}\label{fig4}
\end{figure}
This competition behavior is characterized by the following equation \cite{park,vatansever_3}
\begin{equation}\label{eq9}
\Theta=\frac{t_{1/2}}{\langle \tau\rangle}, 
\end{equation}
where  $\langle \tau\rangle$ is the average relaxation time of the system. 
In order to estimate $\langle \tau\rangle$, we set all spins pointing in the anti-parallel direction 
with respect to a constant bias field $h_{b}/J=0.3$ and monitor the time variation of $m(t)$. In this process, $\langle \tau\rangle$ is defined as the time
at which $m(t)$ momentarily reduces to zero. From our analysis (see Fig. \ref{fig4}), we deduce that  $\langle \tau\rangle=29.6$ on a  DTL which can be compared with
$\langle \tau\rangle=74.6$ of square \cite{robb_1,sides_1,sides_2} and $\langle \tau\rangle=55.8$ of Kagome \cite{vatansever_zd} lattices for the same set of 
other system parameters. Relatively large values obtained in Refs. \cite{robb_1,sides_1,sides_2} are due to the Glauber single-spin-flip algorithm used in the calculations.
It is known that the result will be much smaller for Metropolis dynamics \cite{tauscher}.

\begin{figure}[!h]
\center
\includegraphics[width=7.0cm]{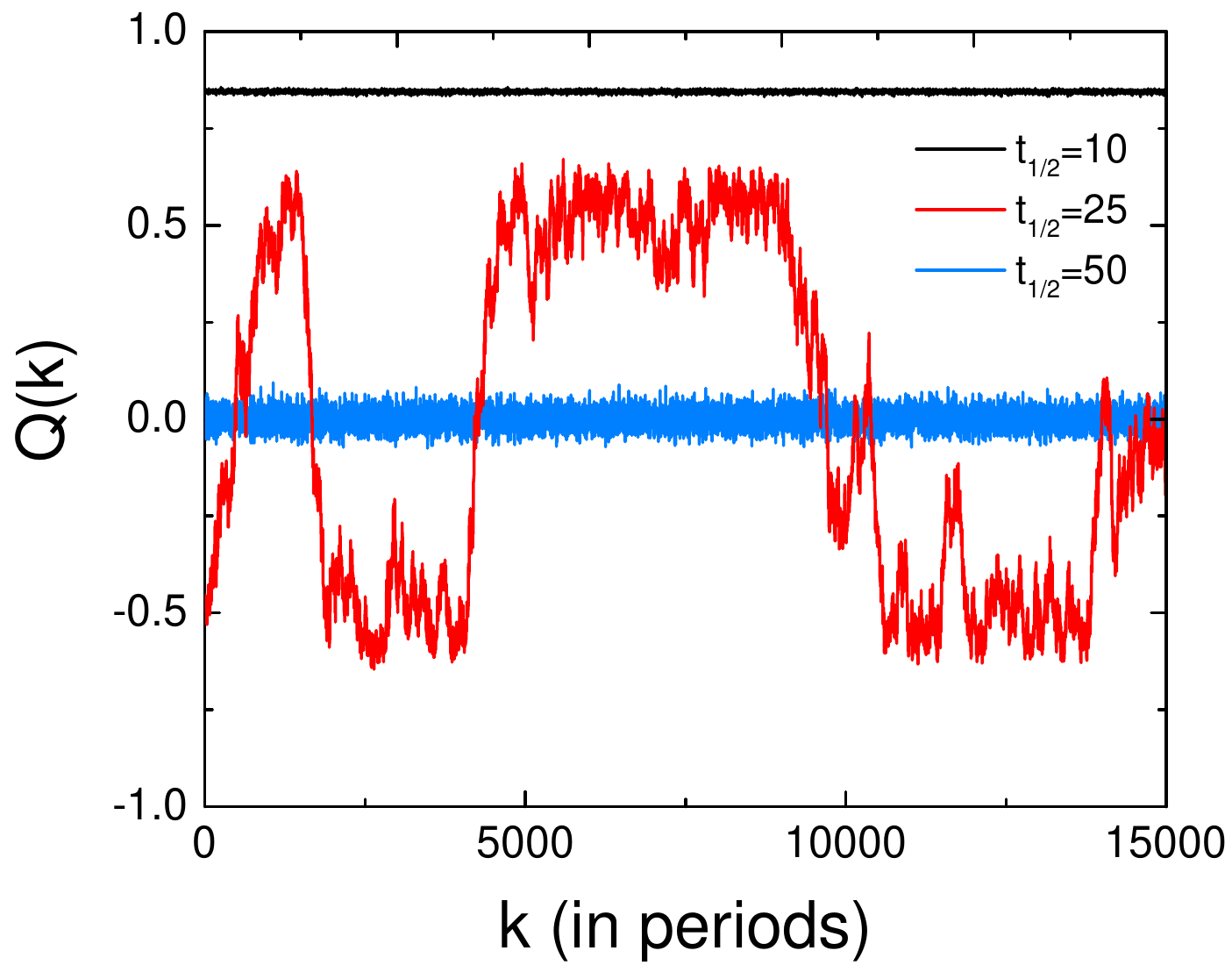}
\caption{Cycle averages of dynamic order parameter $Q$ obtained from the 
time-series of the instantaneous magnetization $m(t)$. 
The system exhibits large fluctuations around the dynamic critical point due to reversal of large 
domains.}\label{fig5}
\end{figure}

The existence of a DPT can be verified by investigating the dynamic order parameter $Q(k)$ calculated by using Eq. (\ref{eq5}) as a function of cycle index $k$ . As shown in Fig. \ref{fig5}, 
below the critical period $(t_{1/2}<t_{1/2}^{c})$, 
a single domain formation is manifested where $Q(k)\neq 0$ whereas for  $(t_{1/2}>t_{1/2}^{c})$, nucleated droplets emerge where $Q(k)\approx0$. 
At the dynamic critical point $\Theta_{c}=t_{1/2}^{c}/\langle \tau\rangle$, a DPT occurs between dynamically ordered and disordered states.

In order to  determine the precise location of the critical half period $t_{1/2}^{c}$ and to estimate the relevant critical exponent ratios, we need to perform finite-size scaling analysis of the numerical data gathered 
in the simulations.
\begin{figure}[!h]
\center
\subfigure[\hspace{0cm}] {\includegraphics[width=7.5cm]{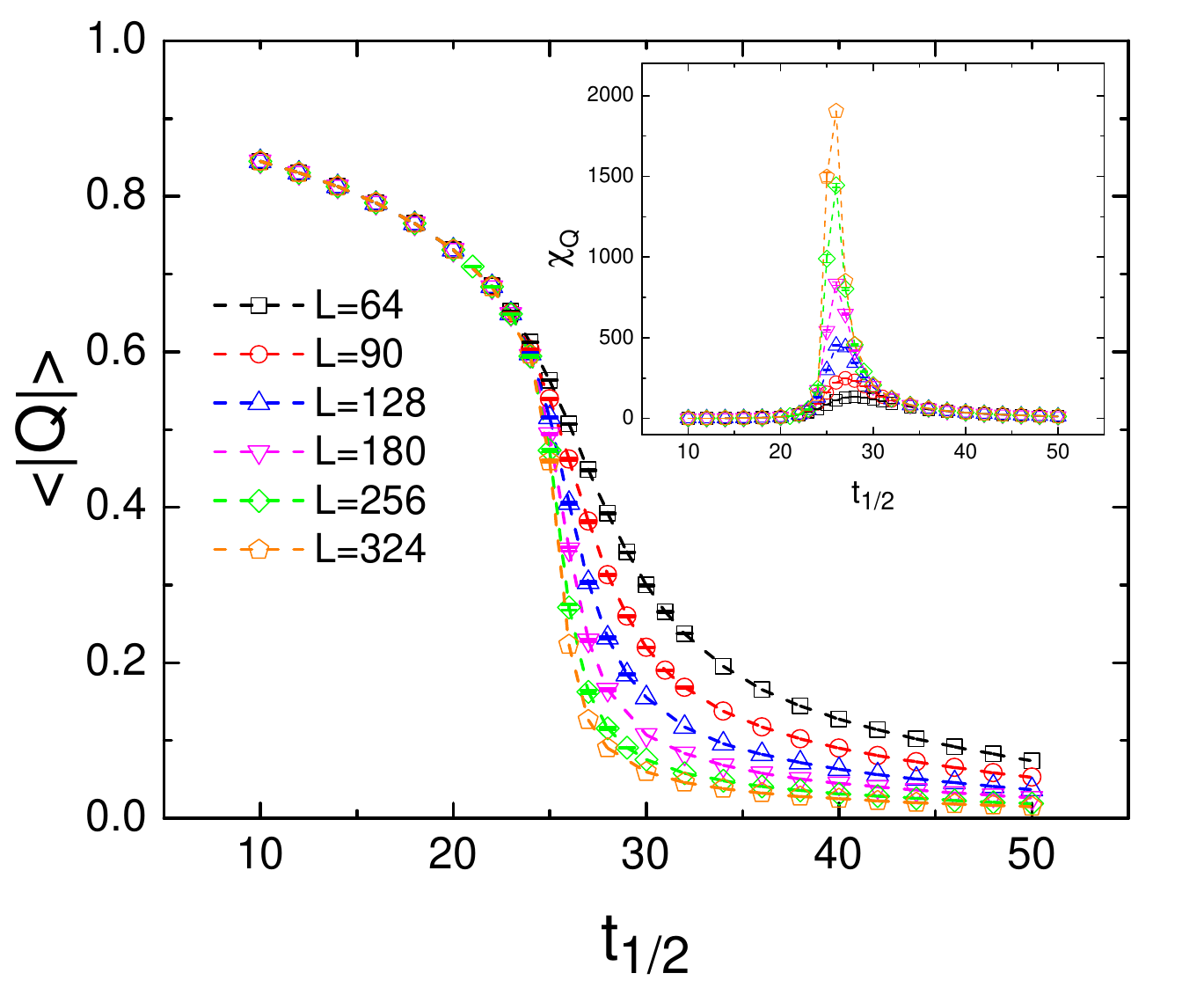}}
\subfigure[\hspace{0cm}] {\includegraphics[width=7.5cm]{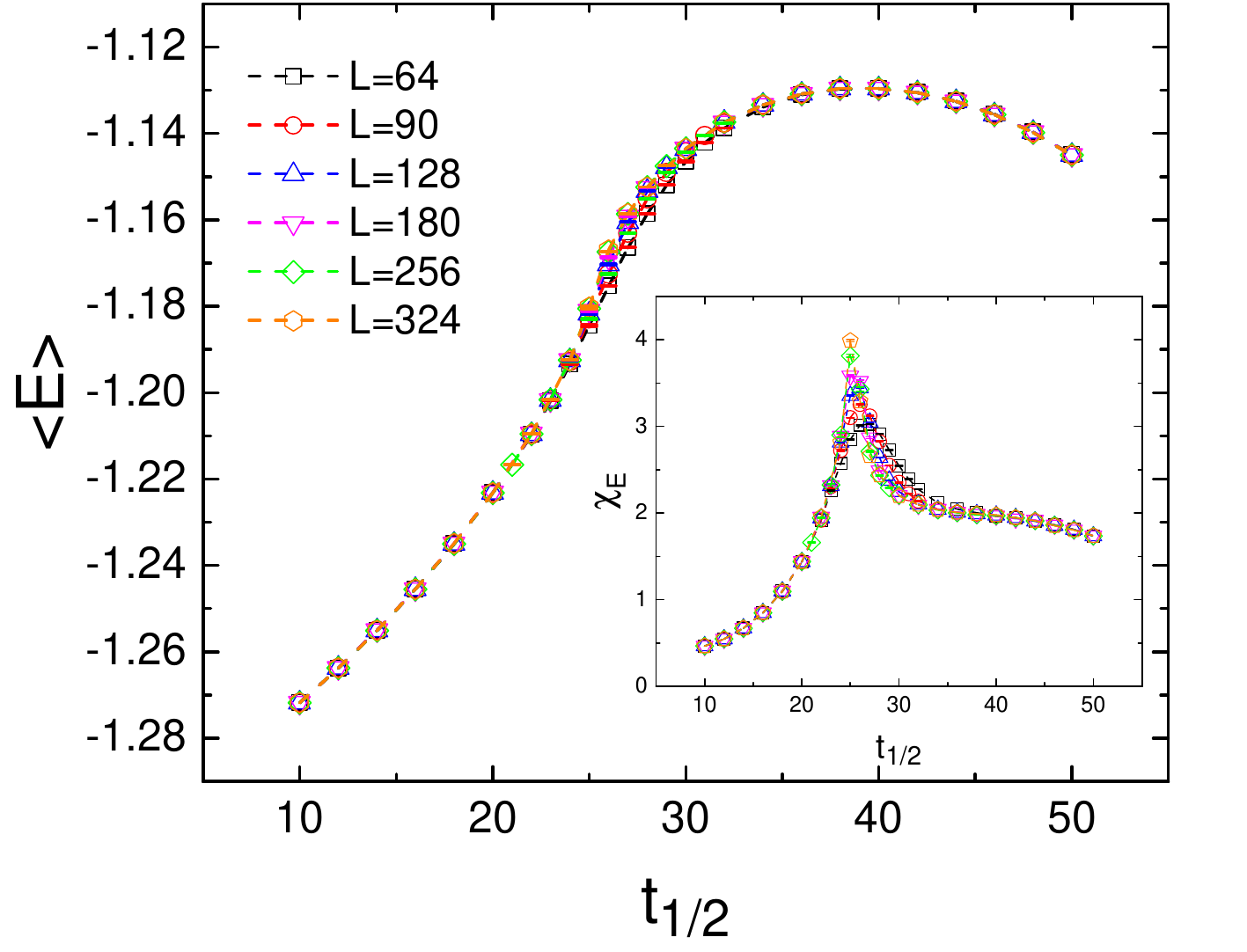}}
\subfigure[\hspace{0cm}] {\includegraphics[width=7.5cm]{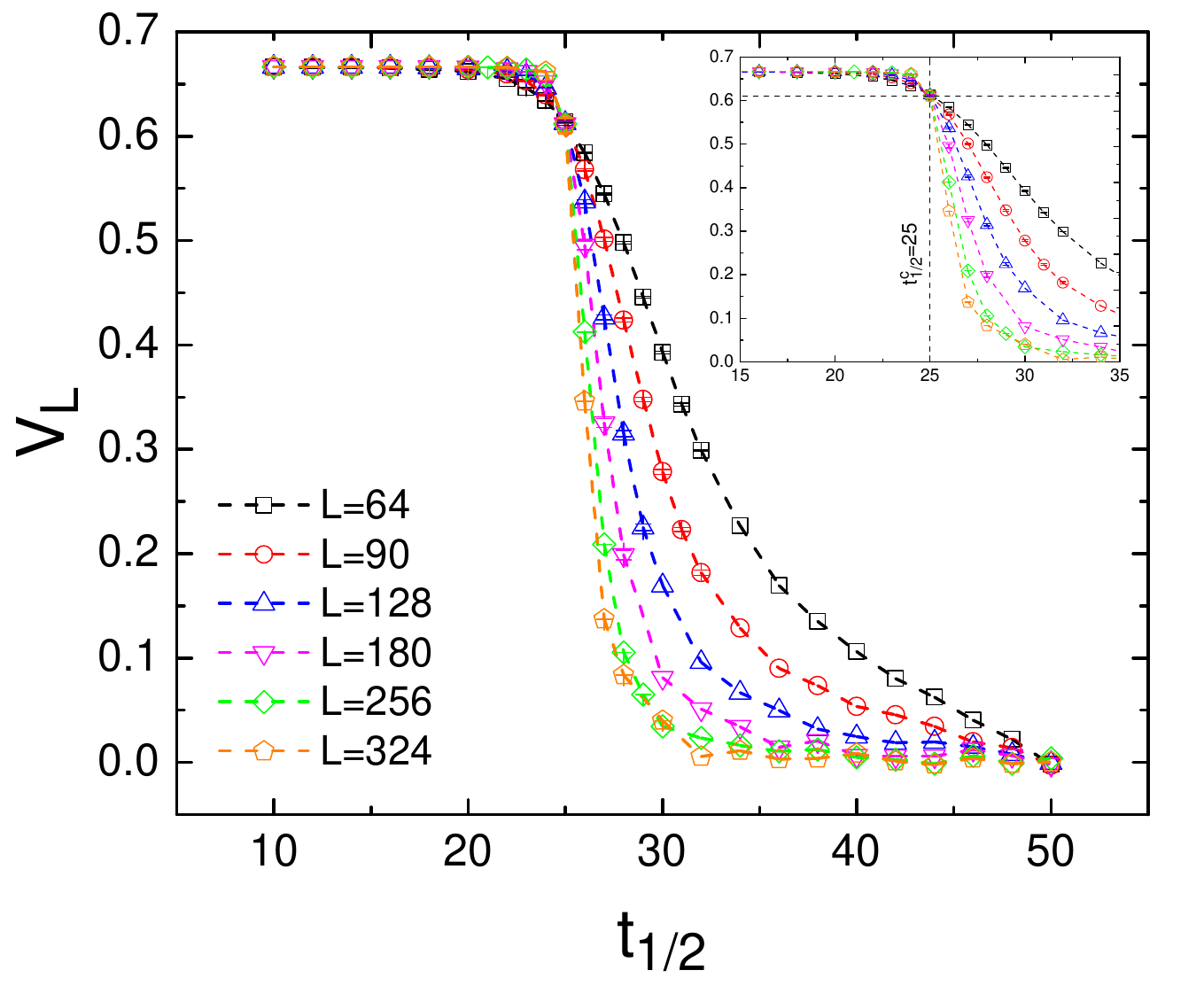}}
\caption{Variation of (a) dynamic order parameter $\langle |Q|\rangle$, (b) 
average energy per-spin $\langle E\rangle$, 
(c) Binder cumulant curve $V_{L}$ as functions of half-period $t_{1/2}$. The 
insets in (a) and (b) respectively correspond to scaled variance curves 
$\chi_{Q}$ and $\chi_{E}$. The inset of Fig. \ref{fig6}c focuses on the critical region.
Different data symbols denote different lattice size $L$.}\label{fig6}
\end{figure}

\subsection*{Finite-size scaling}
The magnetic ordering of the system can be identified by calculating the average of the absolute value of the dynamic order parameter, i.e., $\langle |Q|\rangle$ \cite{buendia}. 
In this regard, $\langle |Q|\rangle$ versus $t_{1/2}$ 
curve of DPT plays the role of the spontaneous magnetization versus temperature curve of TPT. Fig. \ref{fig6}a shows the finite-size behavior of the dynamic order parameter and the corresponding 
response function (scaled variance $\chi_{Q}$). $\langle |Q|\rangle$ decreases from 
its saturation value in the fast critical dynamics regime $(t_{1/2}<t_{1/2}^{c})$ to zero in the slow critical dynamics regime $(t_{1/2}>t_{1/2}^{c})$. The transition is of second-order. 
In the inset of Fig. \ref{fig6}a, we observe that the response function $\chi_{Q}$ exhibits a divergent behavior in the vicinity of the critical point resembling the behavior 
of magnetic susceptibility of a regular ferromagnet. This divergent behavior becomes significant for larger lattices. 
We also calculate the average energy $\langle E\rangle$ and the corresponding scaled variance $\chi_{E}$ as functions of $t_{1/2}$. Both quantities have been plotted in Fig. \ref{fig6}b. 
Note that $\chi_{E}$ mimics the behavior of equilibrium heat capacity which exhibits a prominent cusp around the critical point. 
Very slow variation of $\langle E\rangle$ as a function of system size $L$ around the critical region is clear, and a 
logarithmic scaling behavior of $\chi_{E}$ as a function of $L$ is expected at the dynamic critical point $t_{1/2}^{c}$.

Prior to calculation of critical exponent ratios, we determine the critical half-period $t_{1/2}^{c}$ by measuring the half-period dependence of the fourth-order cumulant (i.e. Binder cumulant) curves according to 
Eq. (\ref{eq8}) for a variety of lattice sizes. In Fig. \ref{fig6}c, the intersection point of the curves is identified as the critical point $t_{1/2}^{c}$. 
According to the simulated data, our estimation is $t_{1/2}^{c}=25$ in units of MCSs. 
The horizontal line (shown in the inset of Fig. \ref{fig6}c) remarks the universal value of the cumulant $V_{L}^{*}=0.6106924(16)$ of the 2D Ising 
model at the critical point \cite{kamieniarz,selke,salas}. This result unveils a similarity between DPT and TPT cases regarding the analyses of Binder cumulant curves. 

\begin{figure}[!h]
\center
\subfigure[\hspace{0cm}] {\includegraphics[width=7.5cm]{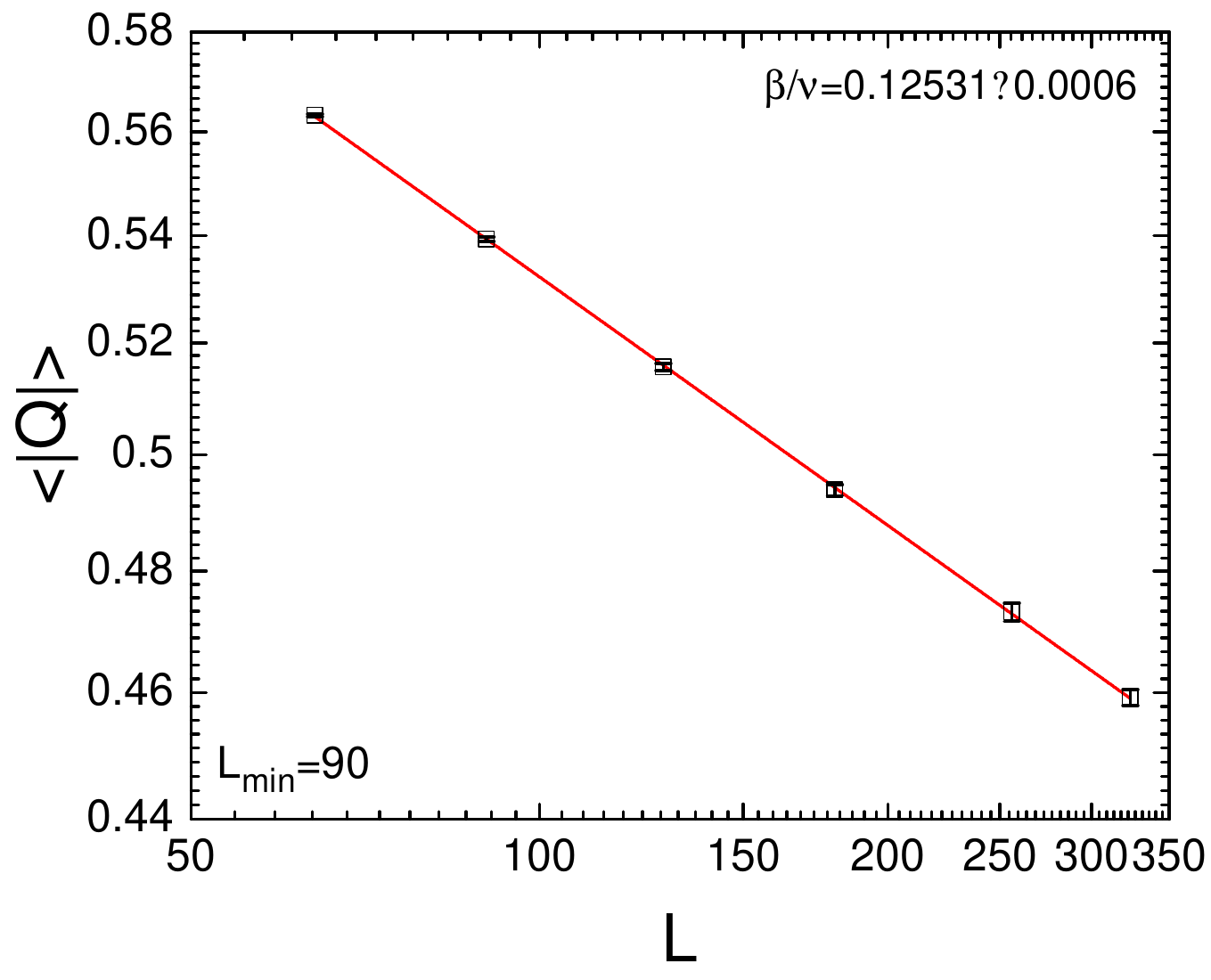}}
\subfigure[\hspace{0cm}] {\includegraphics[width=7.5cm]{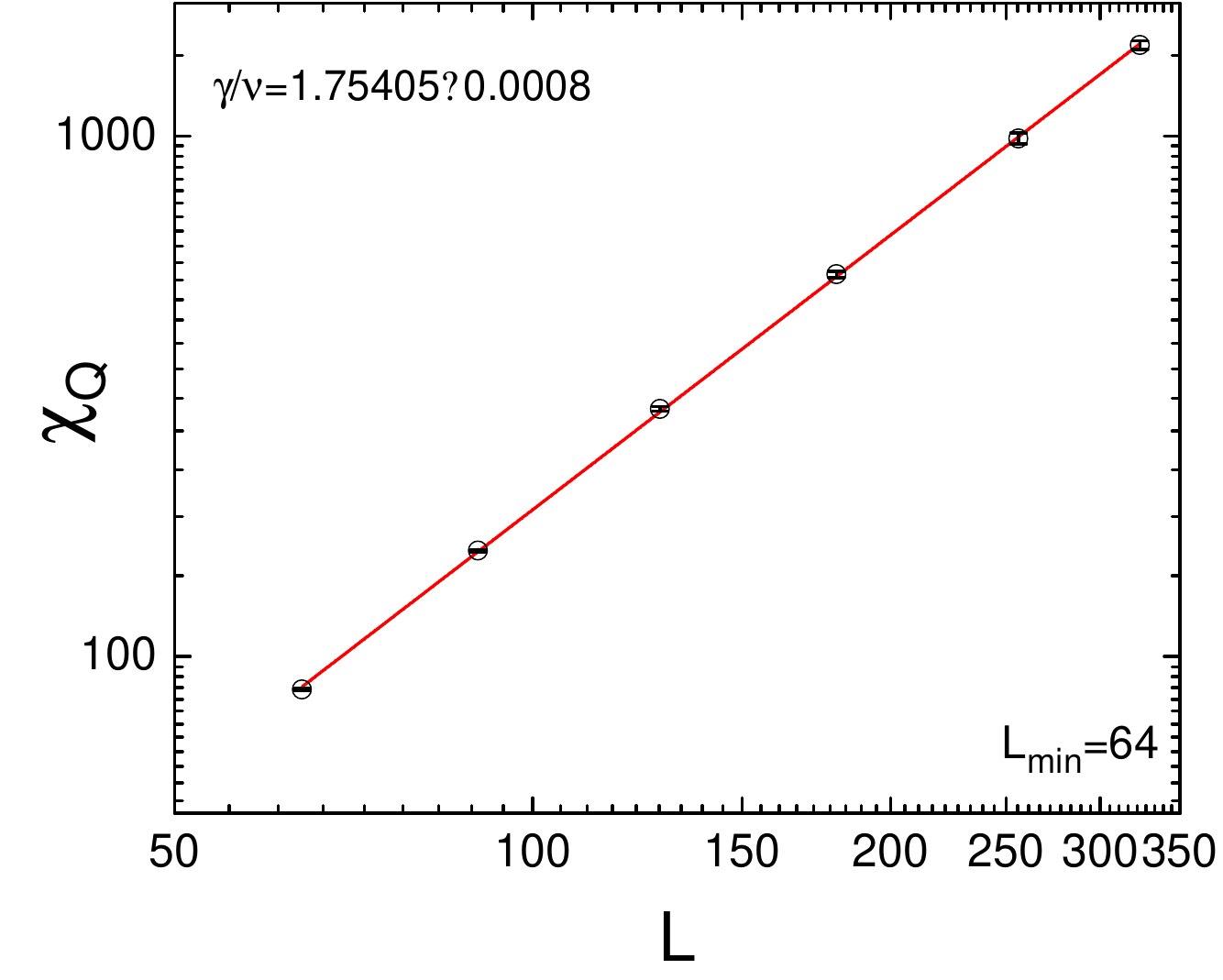}}
\subfigure[\hspace{0cm}] {\includegraphics[width=7.5cm]{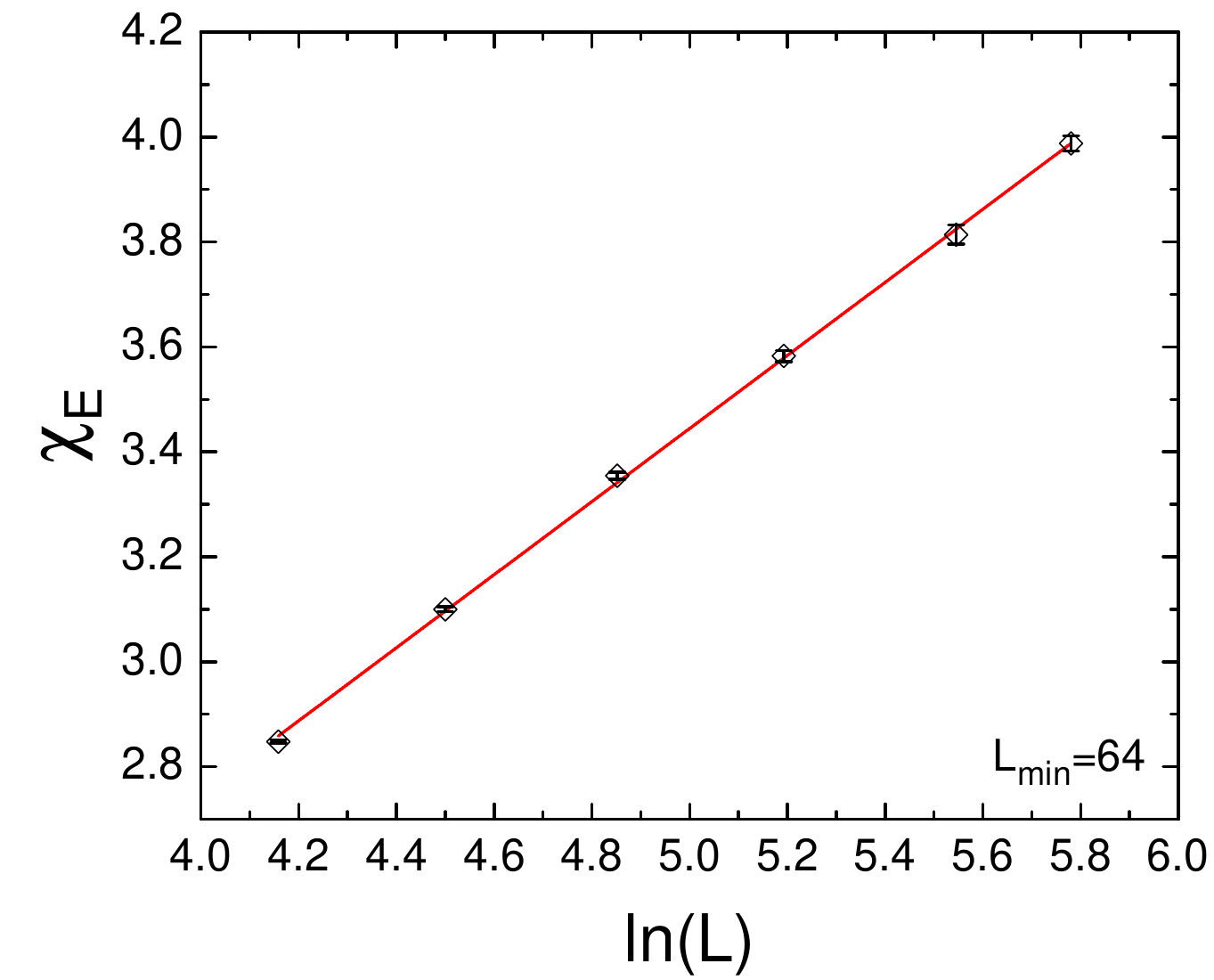}}
\caption{Finite-size scaling analysis of (a) $\langle |Q|\rangle$, (b) $\chi_{Q}$, (c) $\chi_{E}$. 
Solid red lines represent the linear fitting curves.}\label{fig7}
\end{figure}
At the dynamic critical point, dynamic order parameter  $Q$  and the scaled variance $\chi_{Q}$ obey the following scaling forms \cite{buendia,vatansever_e,park,ev_fytas}
\begin{eqnarray}\label{eq10}
\langle |Q|\rangle&\propto&  L^{-\beta/\nu},\\
\chi_{Q}&\propto&L^{\gamma/\nu}.
\end{eqnarray}
Logarithmic plots of $\langle |Q|\rangle$ and $\chi_{Q}$ as functions of $L$ obtained at the critical point $t_{1/2}^{c}$ 
exhibit a linear variation. After fitting the data, we find $\beta/\nu=0.12531\pm0.0006$ and $\gamma/\nu=1.75405\pm0.000795$ (Figs. \ref{fig7}a,b)
which agree well with the 2D Ising equilibrium results $\beta/\nu=1/8$ and $\gamma/\nu=7/4$ within the estimated errors \cite{binder,fisher}. 

Besides, the logarithmic divergence behavior of $\chi_{E}$ at $t_{1/2}^{c}$
\begin{equation}\label{eq11}
\chi_{E}\propto a+b\ln(L), 
\end{equation}
can be observed in semi-logarithmic plot of $\chi_{L}^{E}$ as a function of $L$, indicating that the  related exponent has the value $\alpha=0.0$ (Fig. \ref{fig7}c). 

Consequently, these results hitherto show that the estimated critical exponents agree well with the previous results \cite{yuksel1}, 
and it can be once again emphasized that the DPT falls within the same universality class as the TPT.

\subsection*{Metamagnetic anomalies}
\begin{figure}[!h]
\center
\subfigure[\hspace{0cm}] {\includegraphics[width=4.0cm]{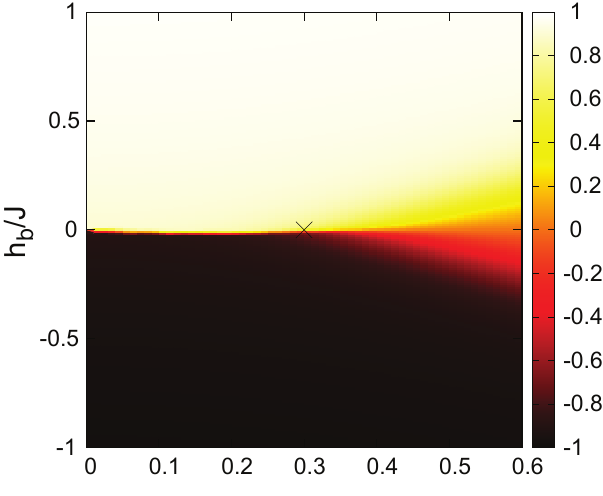}}
\subfigure[\hspace{0cm}] {\includegraphics[width=3.85cm]{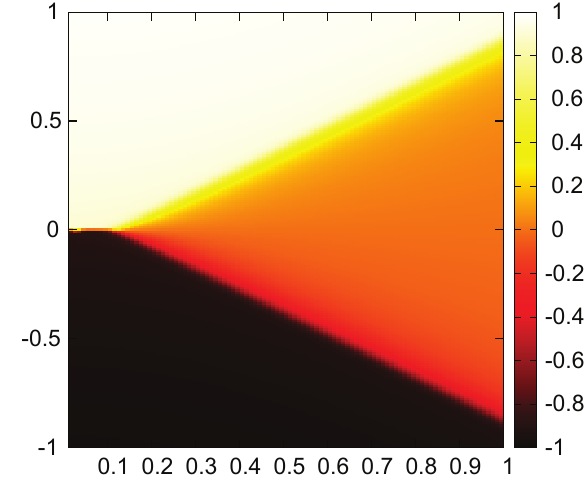}}\\
\subfigure[\hspace{0cm}] {\includegraphics[width=4.0cm]{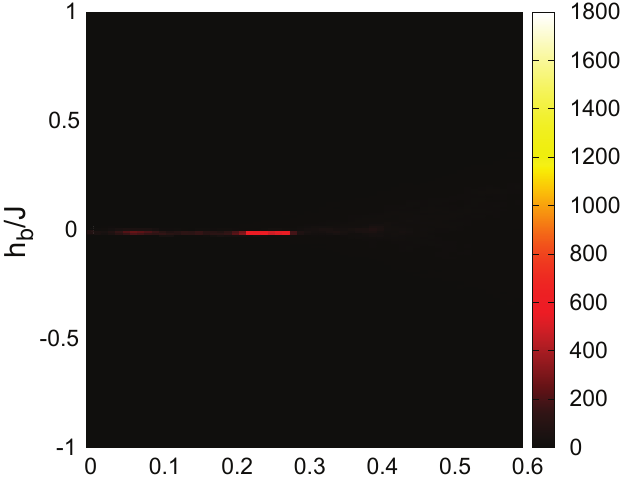}}
\subfigure[\hspace{0cm}] {\includegraphics[width=3.85cm]{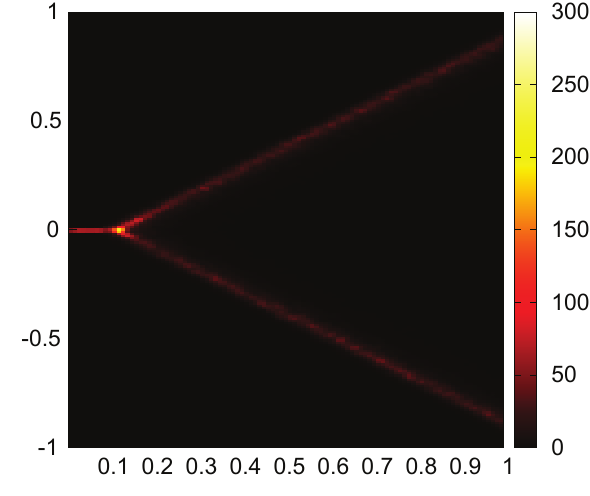}}\\
\subfigure[\hspace{0cm}] {\includegraphics[width=4.0cm]{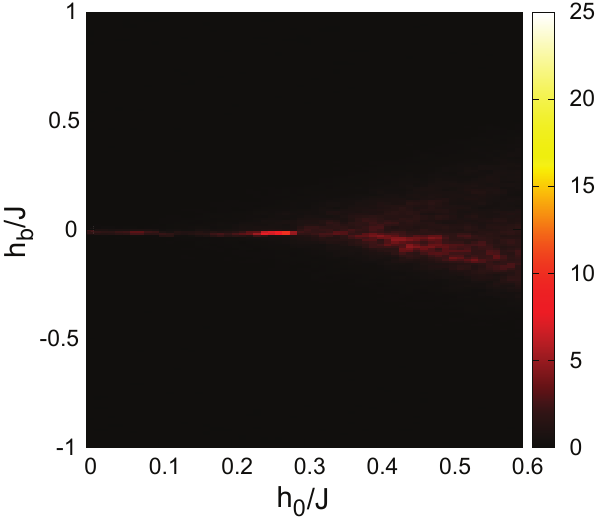}}
\subfigure[\hspace{0cm}] {\includegraphics[width=3.85cm]{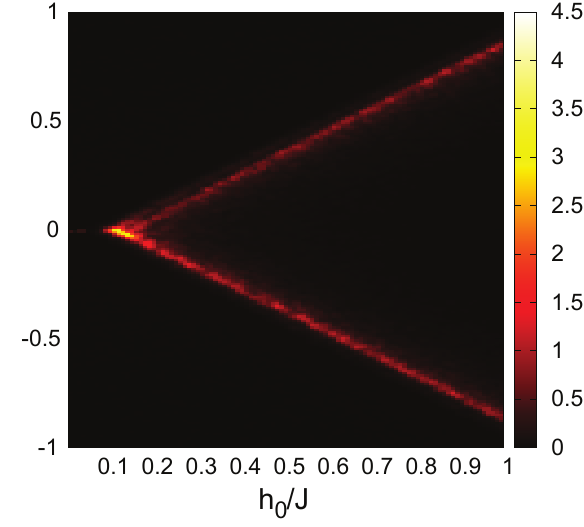}}\\
\caption{Contour plots of (a), (b): dynamic order parameter $\langle Q\rangle$; 
(c), (d): scaled variance $\chi_{Q}$; (e), (f): scaled variance $\chi_{E}$ in 
$h_{0}/J$ versus $h_{b}/J$ plane. The left panel has been obtained for 
a field period $t_{1/2}=25$ whereas for the right panel we set $t_{1/2}=250$. Data point $\times$ in (a) denote the dynamic critical point.}\label{fig8}
\end{figure}

So far, we  have elucidated some salient similarities between DPT and TPT cases. These similarities mainly originate in the vicinity of critical point, and in the absence of bias field $h_{b}$. 
In addition to the periodically oscillating part, upon introducing 
a time-independent contribution $h_{b}$ in the magnetic field term, some controversial behaviors can be observed. For example, a figure of merit for the aforementioned issue is the metamagnetic anomaly phenomenon which is 
especially observed  in the slow critical dynamics regime. It was experimentally reported for Co films that 
the metamagnetic anomaly behavior is manifested in the $h_{b}$ dependence of $\chi_{Q}$ as multiple-symmetric peaks \cite{riego}, despite the fact that 
it is not observed for a regular ferromagnet for which the magnetic susceptibility versus magnetic field curve exhibits a broad maxima centered around zero field \cite{berger_eq}.
Recently, the experimental observations of Refs. \cite{riego,ramirez} have been supported by some theoretical studies \cite{buendia2,shi,yuksel3,yuksel2}. 
In the following, we present our simulation results for the present model. In Fig. \ref{fig8}, we  show the contour plots of the quantities $\langle Q\rangle$, $\chi_{Q}$ and $\chi_{E}$ as 
functions of the field parameters $h_{0}/J$ and $h_{b}/J$. The left panel of Fig. \ref{fig8} shows the results obtained at $t_{1/2}^{c}=25$ whereas for the right panel we set $t_{1/2}=250>>t_{1/2}^{c}$. 
Below the dynamic critical point which is marked by the symbol ``$\times$'' in Fig. \ref{fig8}a, 
$(\langle Q\rangle,h_{0}/J,h_{b}/J)$ plots exhibit discontinuous jumps between $\langle Q\rangle=\pm Q_{0}$ values. 
By comparing Figs. \ref{fig8}a and \ref{fig8}b with each other, we see that critical amplitude value shifts to smaller values for increasing field period which means that the dynamic paramagnetic region in the phase space
becomes expanded.
This behavior results in as the triangular regions depicted in Figs. \ref{fig8}a and \ref{fig8}b. 
A steep variation of $\langle Q\rangle$ around $h_{b}=h_{b}^{peak}$ is evident
whereas for $h_{b}>h_{b}^{peak}$,  $\langle Q\rangle$ saturates to unity. Although the metamagnetic anomalies (i.e. the side bands) are not visible at the dynamic critical point $(h_{0}/J=0.3,t_{1/2}^{c}=25)$ 
in $\chi_{Q}$ contour plot (Fig. \ref{fig8}c), the phenomenon is pronounced in the slow critical dynamics regime (Fig. \ref{fig8}d).  
Moreover, $h_{b}$ values corresponding to $\chi_{Q}$ peak positions 
represent a critical threshold 
indicating that for $h_{b}>h_{b}^{peak}$ we observe dynamically ferromagnetic (polarized) oscillations. On the other hand, for $h_{b}<h_{b}^{peak}$ the system stays in the dynamically paramagnetic phase.
We have also examined the emergence of metamagnetic anomalies in $\chi_{E}$ contour plots. As we found for $\chi_{Q}$ curves, the side-band behavior is barely evident in the vicinity of dynamic critical point whereas
they are indisputably pronounced in the slow critical dynamics regime where $t_{1/2}>>t_{1/2}^{c}$. These results clearly suggest that, in regard to dissimilarities between DPT and TPT cases, 
beside the scaled variance $\chi_{Q}$, the other response function $\chi_{E}$ also exhibits metamagnetic anomalies which are very prominent in the slow critical dynamics regime. Such behavior has also not been 
observed in equilibrium ferromagnets. 

\begin{figure}[!h]
\center
\includegraphics[width=7.0cm]{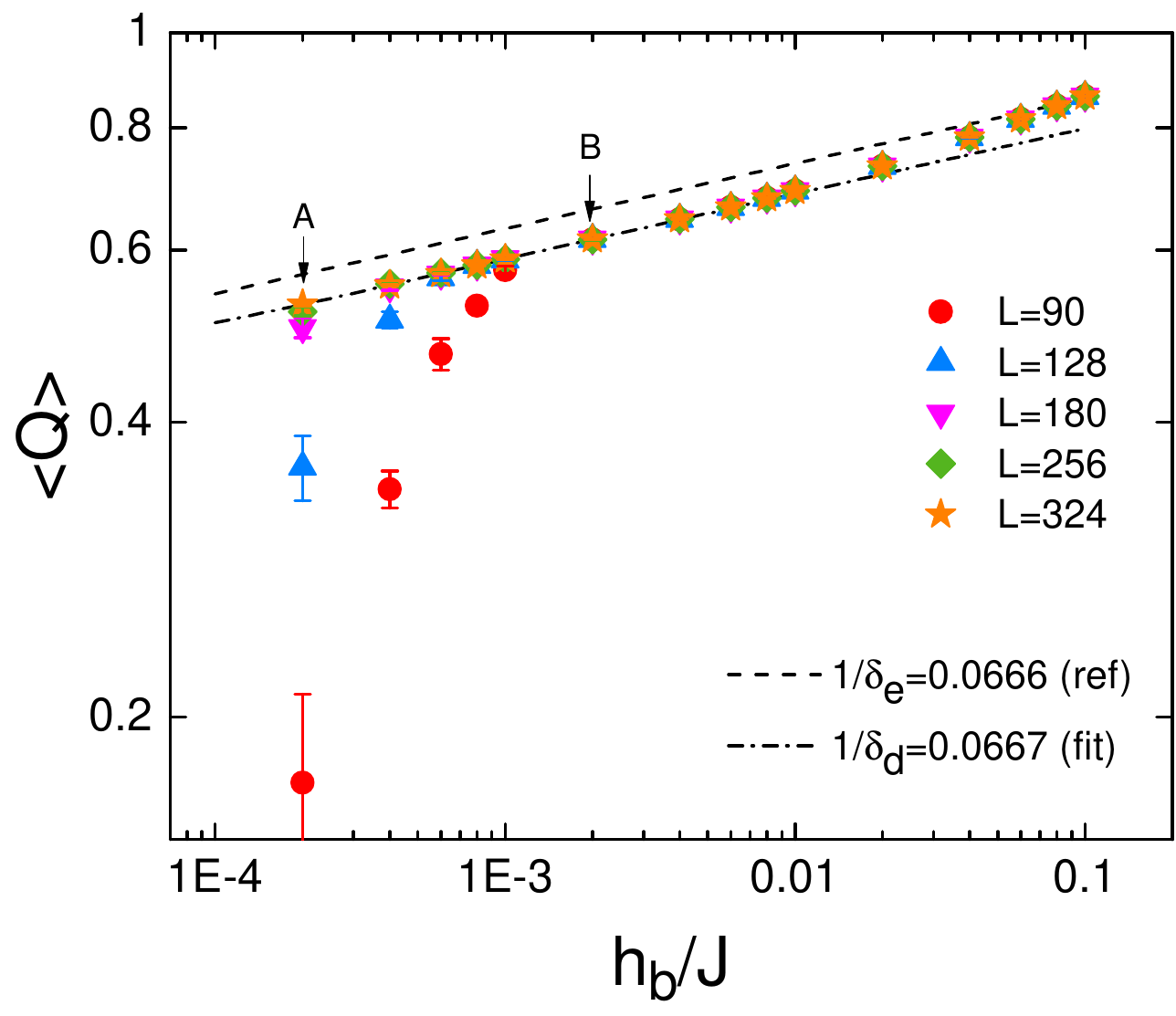}
\caption{Dependence of $\langle Q\rangle$ as a function of bias field $h_{b}$ depicted in the log-log scale for a variety of lattice sizes within the 
range $90\leq L\leq 324$. The dotted line is the result for $L=324$ where a linear fitting procedure was performed between the labels $\mathrm{A}$ and $\mathrm{B}$. 
The dashed line corresponds to the scaling of the equilibrium model \cite{mcKenzie}.}\label{fig9}
\end{figure}
Last but not least, another DPT critical exponent value can be found related to the scaling of $\langle Q\rangle$ with respect to $h_{b}$ in the form
\begin{equation}\label{eq13}
\langle Q\rangle(t_{1/2}=t_{1/2}^{c},h_{b}\rightarrow0)\propto h_{b}^{1/\delta_{d}}, 
\end{equation}
where a dynamical scaling exponent $\delta_{d}=14.85$ was estimated for a square lattice under the influence of a square wave field within the small $h_{b}$ regime \cite{robb_1}.
This result is very close to the critical isotherm value $\delta_{e}=15$ of the 2D Ising model in equilibrium \cite{mcKenzie}. For a DTL, we perform simulations in the vanishingly small $h_{b}$ regime 
for lattice sizes ranging from $L=90$ to $L=324$ for $t_{1/2}=25$ and $h_{0}/J=0.3$ (Fig. \ref{fig9}) . In case of large $L$ and small $h_{b}$, log-log plot of $(\langle Q\rangle \ \mathrm{vs} \ h_{b})$
plots show a linear behavior, and considering the  numerical data corresponding to $L=324$ between the points $\mathrm{A}$ and  $\mathrm{B}$, 
the extracted exponent value is found to be $\delta_{d}=14.99$ which improves Ref. \cite{robb_1}. This observation  also proves the fact that the bias field  $h_{b}$ is the conjugate field of the dynamic order parameter $Q$.    

\section{Conclusion}\label{conclude}

In summary, we perform extensive Monte Carlo simulations to explore the phase transition characteristics and critical behavior of 2D Ising model on a decorated triangular lattice. In the first part of the work, 
we identify the critical temperature of the model in the absence of dynamic magnetic field effects (equilibrium model).
The model exhibits a ferromagnetic-paramagnetic phase transition at a critical temperature $T_{c}/J=1.75$ which is smaller than the result corresponding to regular triangular lattice. This is attributed to the small 
effective coordination number of the DTL in comparison to its regular counterpart.

The second part of the study is focused on the DPT properties. In this regard, 
using the finite-size scaling theory and its arguments, dynamic critical point of the lattice is found to be $t_{1/2}^{c}=25$ in units of MCSs, with the  respective estimations of the critical 
exponent ratios $\beta/\nu=0.125$ and $\gamma/\nu=1.754$ 
associated to the dynamic order parameter $\langle Q\rangle$ and the corresponding response function $\chi_{Q}$, 
leading to the fact that DPT falls in the same universality class as the TPT, supporting the observations of the previous works. 
Moreover, a logarithmic scaling behavior in the scaled variance $\chi_{E}$ as a function of $L$ is also predicted at the critical point $t_{1/2}^{c}$.	 
These results  indicate that DPT and TPT properties show a number of similarities in the vicinity of critical point and in the absence of magnetic bias field $h_{b}$.

Upon introduction of the non-zero bias field, some peculiarities called ``metamagnetic anomalies`` 
originate in the magnetic behavior of the system  which cannot be observed in the equilibrium case. 
Our results confirm that this side-band phenomenon observed in $(\chi_{Q},h_{0}/J,h_{b}/J)$ contour plots also emerge for  $(\chi_{E},h_{0}/J,h_{b}/J)$ contours
which become very prominent in the slow critical dynamics regime. Nevertheless, another similarity between DPT and TPT cases can be captured when we consider log-log plots of 
$(\langle Q\rangle \ \mathrm{vs} \ h_{b})$ curves for small $h_{b}$ regime at the dynamic critical point $t_{1/2}^{c}$,   
resulting in another critical exponent $\delta_{d}=14.99$ for large $L$ which is very close to the critical isotherm of the equilibrium model. 
Taking into account this latter issue, we can conclude that our new result improves the findings of previously published works.  

Overall, although the critical behavior of equilibrium magnetic systems have been well established, the theory of dynamic phase transitions tends to flourish within the last two decades.
As an outlook, we hope that our results reported in this paper would make a contribution 
on the pursue of new concepts in dynamic phase transitions, and would stimulate further studies for the investigation of 2D magnetism, both from theoretical and experimental points of view.
As a final remark, in order to clarify the role of the lattice structure on the DPT properties of spin systems with increased complexity,
the problem handled in the present work can be extended to some other forms of irregular lattices including 
decorated square, and decorated simple cubic lattices with more complex Hamiltonian forms including Blume-Capel and Blume-Emery-Griffiths models. 
Simulations in this regard are under consideration by us, and it may be the subject of a future work. 
\section*{Appendix}
\begin{figure}[!h]
\center
\subfigure[\hspace{0cm}] {\includegraphics[width=4.0cm]{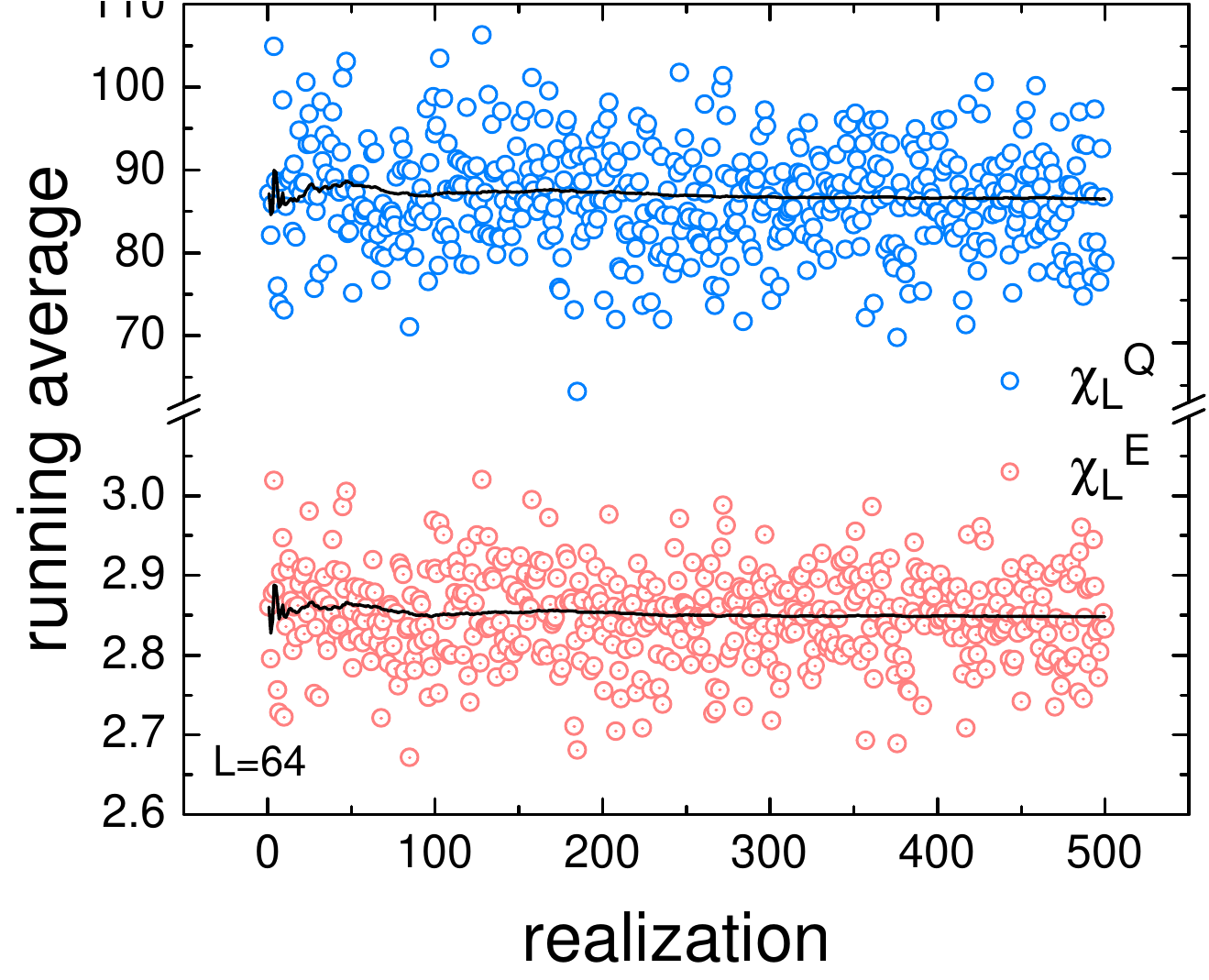}}
\subfigure[\hspace{0cm}] {\includegraphics[width=4.0cm]{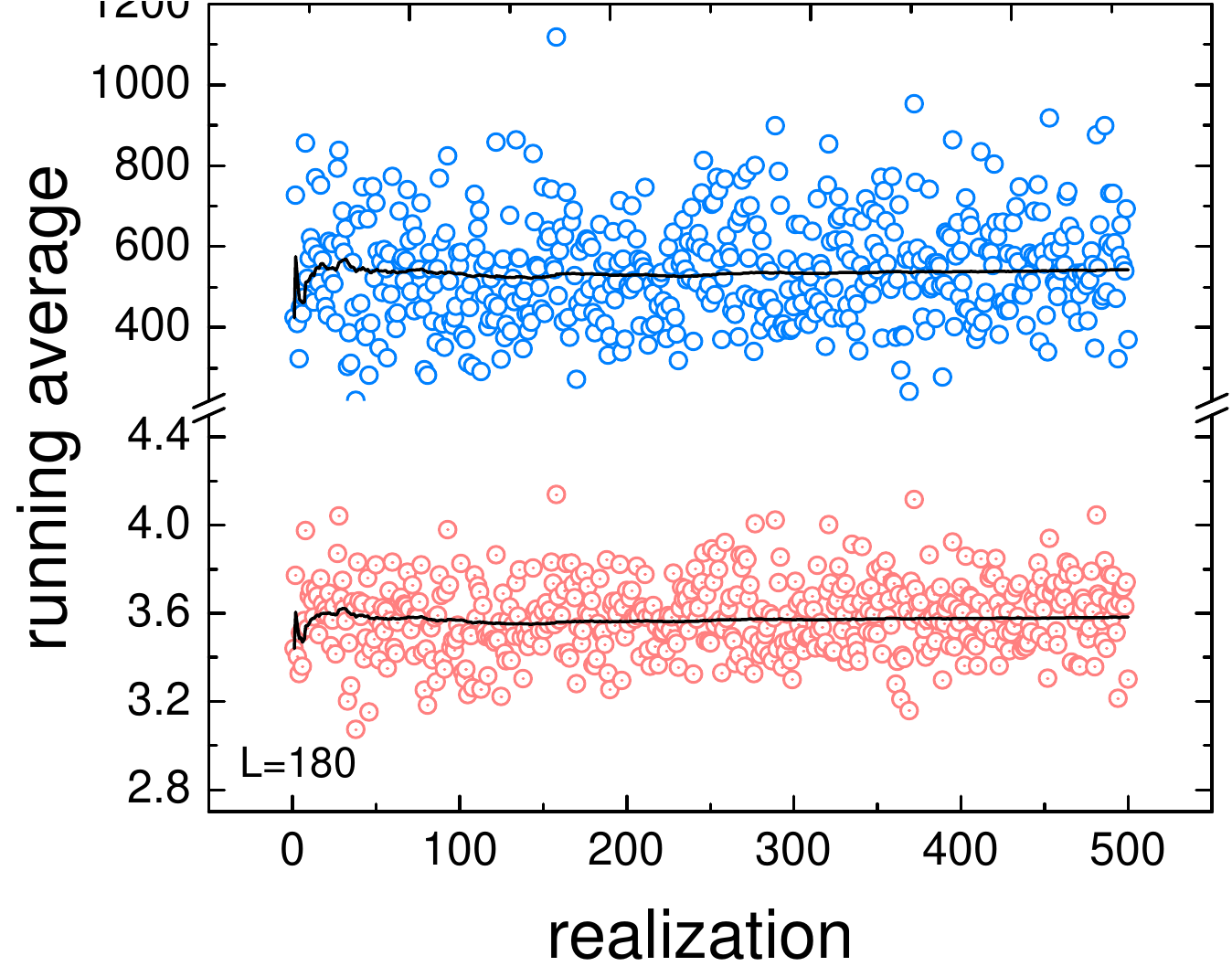}}\\
\subfigure[\hspace{0cm}] {\includegraphics[width=4.0cm]{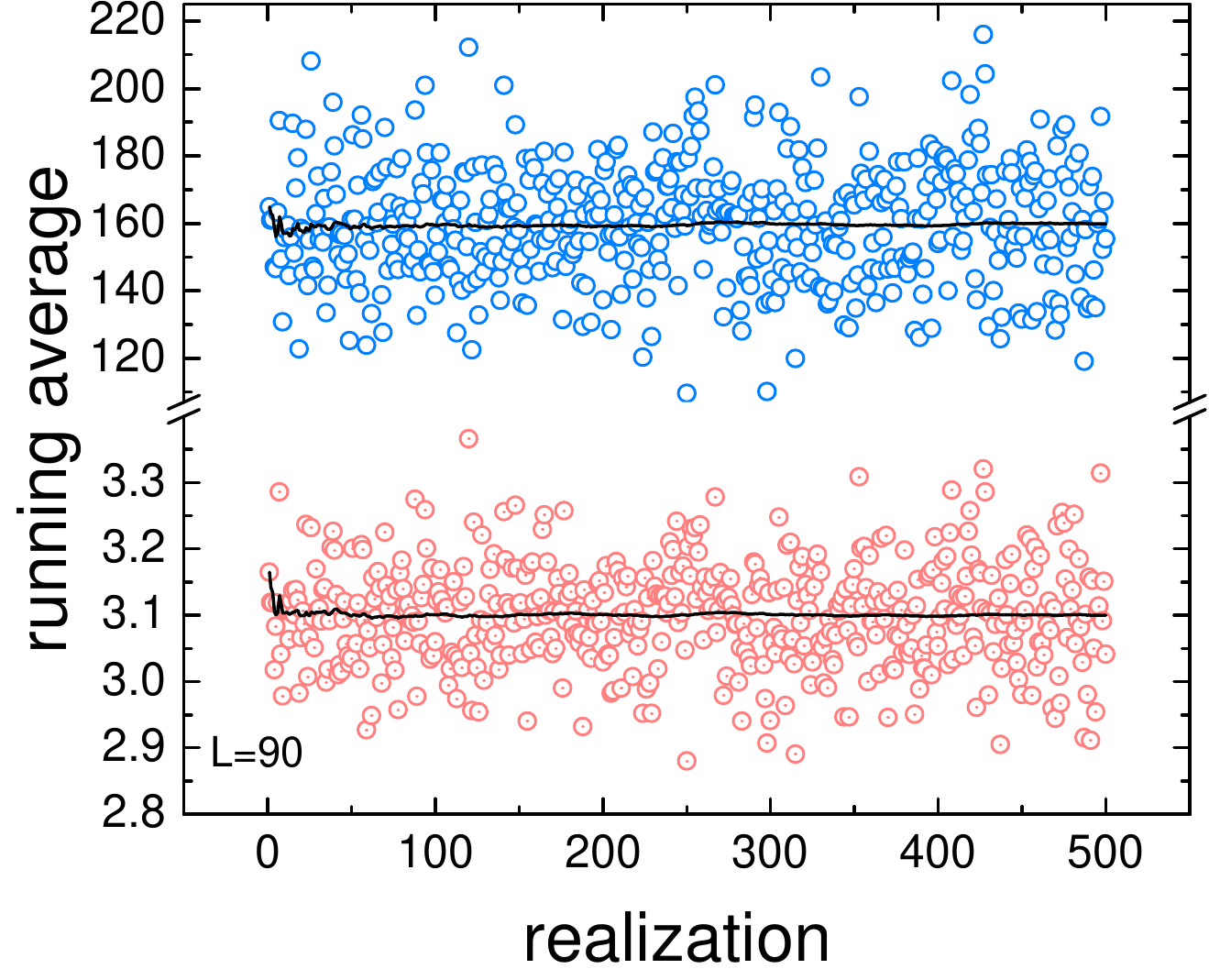}}
\subfigure[\hspace{0cm}] {\includegraphics[width=4.0cm]{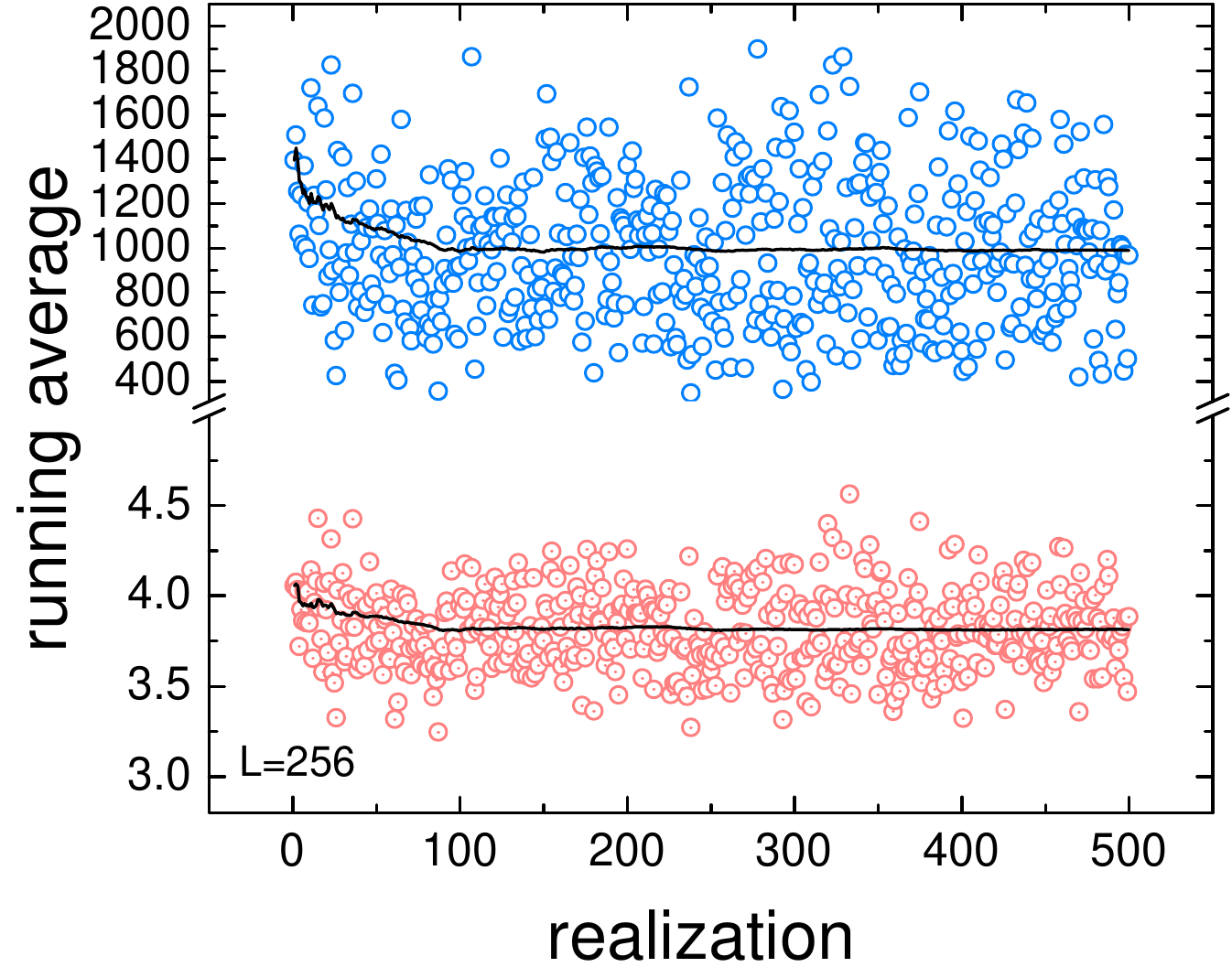}}\\
\subfigure[\hspace{0cm}] {\includegraphics[width=4.0cm]{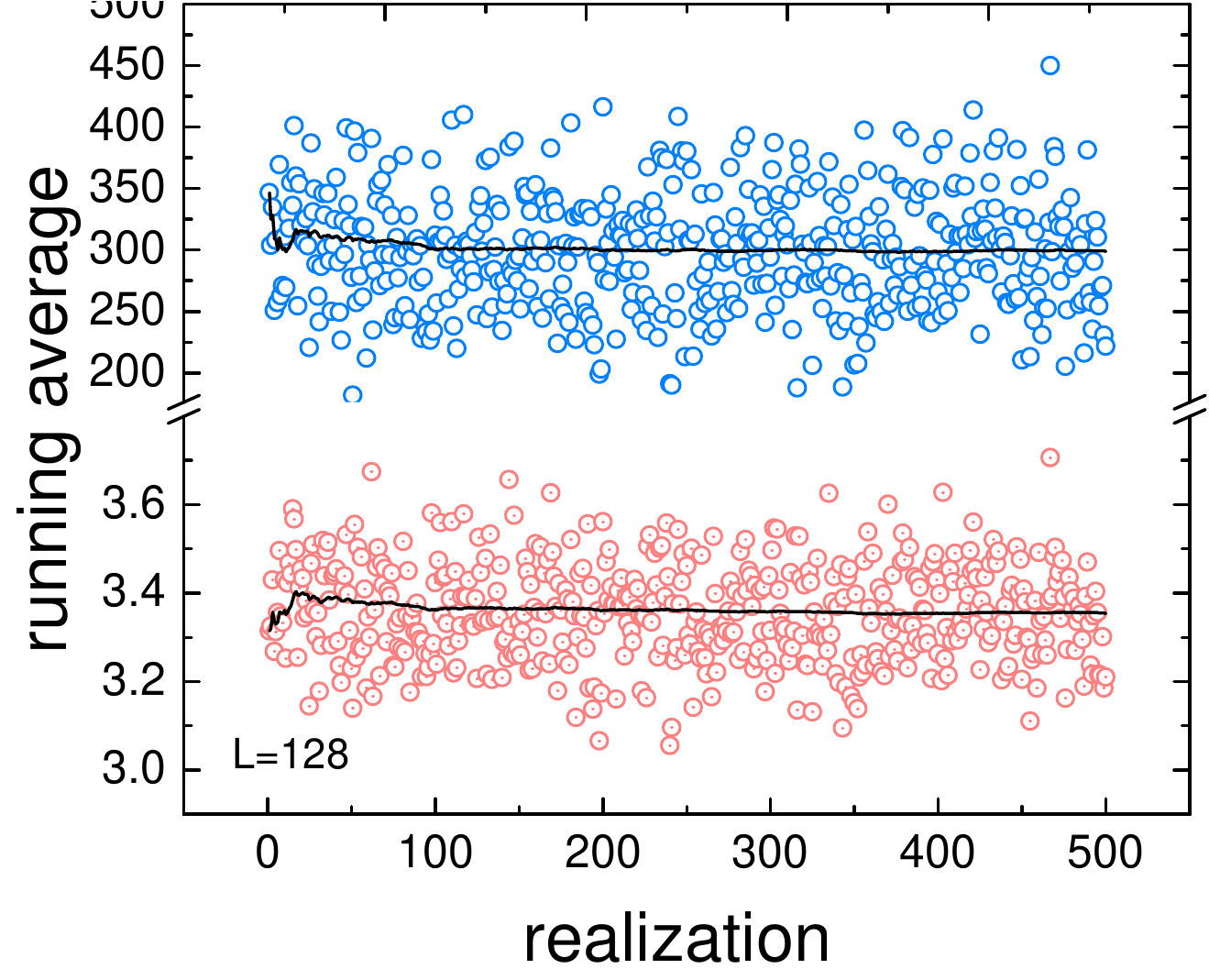}}
\subfigure[\hspace{0cm}] {\includegraphics[width=4.0cm]{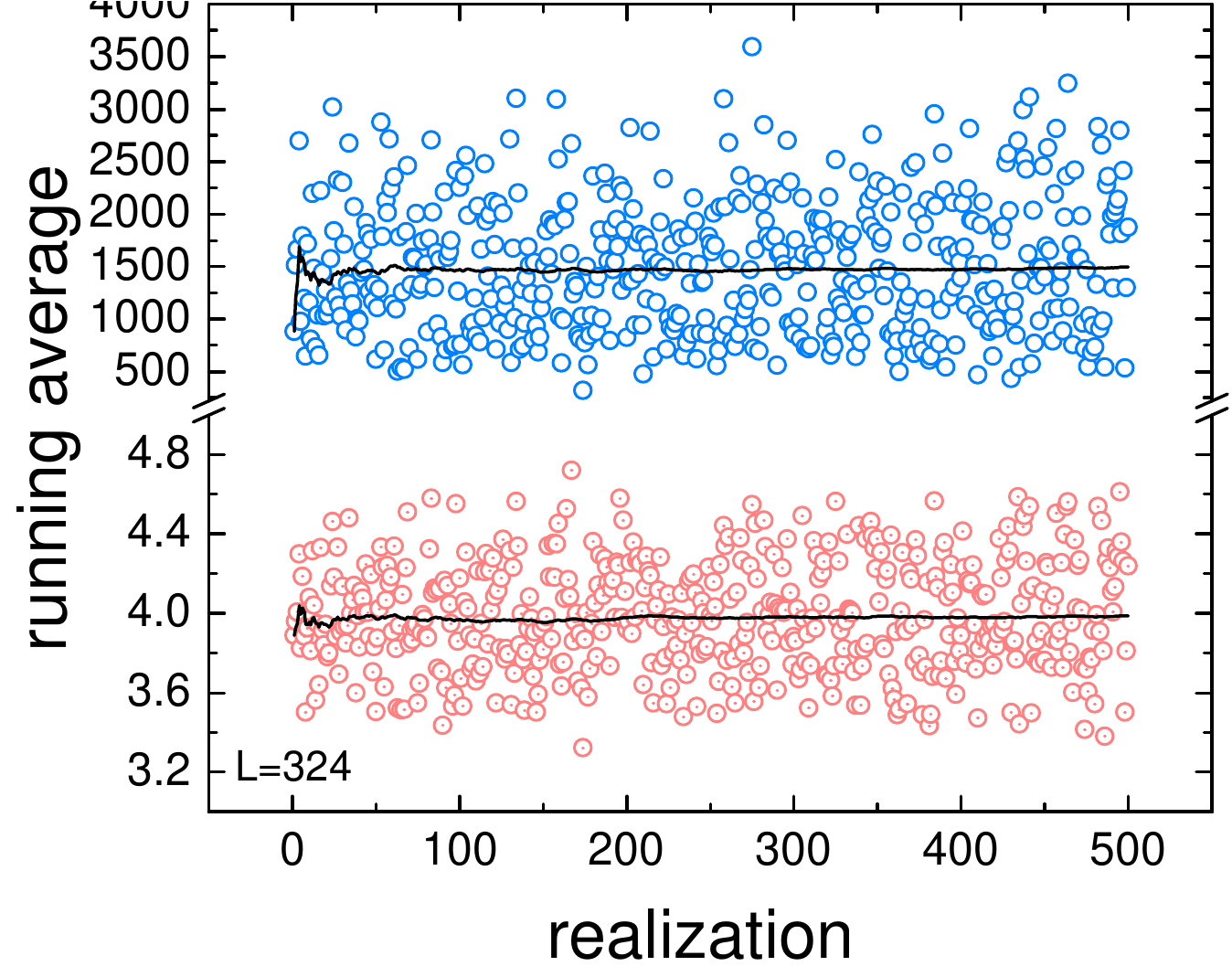}}\\
\caption{Running averages of $\chi_{L}^{Q}$ and $\chi_{L}^{E}$ calculated at dynamic critical point $t_{1/2}^{c}=25$ over 
500 independent realizations . 
Data points represent individual measurements. Solid lines denote the 
accumulated averages of the aforementioned quantities.}\label{fig10}
\end{figure}

A test of the running averages for the zero-field response functions $\chi_{Q}$ and $\chi_{E}$ have been performed for $500$ independent sample realizations, and the results are given in Fig. \ref{fig10}. 
Figure shows that the accumulated averages (solid curves) of the aforementioned quantities saturate at a fixed point for $64\leq L\leq 324$ indicating that $500$ independent samples may be sufficient to reduce 
the statistical errors regarding the estimation of dynamic critical point and critical exponent ratios. 

\begin{acknowledgments}
The computational resources are provided by TUBITAK ULAKBIM, High Performance 
and Grid Computing Center (TR-Grid e-Infrastructure).
\end{acknowledgments}

\bibliography{references}

\end{document}